\RequirePackage{fix-cm}
\documentclass[twocolumn,epjc3]{svjour3}  

\pdfoutput=1

\smartqed  
\RequirePackage{graphicx}
\RequirePackage{widetext}
\RequirePackage{float}
\RequirePackage{subfigure}
\RequirePackage{amssymb}   
\RequirePackage{amsmath}   
\RequirePackage{epsfig}

\def\la{\left(}
\def\rf{\right)}
\def\|{\'\i}
\def\la{\left(}
\def\rf{\right)}

\def\ed{\end{document}}
\def\be{\begin{equation*}}
\def\ee{\end{equation*}}
\def\beq{\begin{eqnarray}}
\def\eq{\end{eqnarray}}

\RequirePackage[colorlinks,citecolor=blue,urlcolor=blue,linkcolor=blue]{hyperref}
\journalname{Eur. Phys. J. C}

\begin{document}

\title{Symmetry breaking patterns of the 3-3-1 model at finite temperature}

\author{J. S\'a Borges\thanksref{e1,addr1}
\and
Rudnei O. Ramos\thanksref{e2,addr2}
}

\thankstext{e1}{e-mail: saborges@uerj.br}
\thankstext{e2}{e-mail: rudnei@uerj.br}

\institute{
Departamento de F\'{\i}sica de Altas Energias,
  Universidade do Estado do Rio de Janeiro, 20550-013 Rio de Janeiro,
  RJ, Brazil \label{addr1} 
\and 
Departamento de F\'{\i}sica Te\'orica, 
Universidade do Estado do Rio de Janeiro, 20550-013 Rio
de Janeiro, RJ, Brazil \label{addr2}
}

\date{Received: date / Accepted: date}

\maketitle

\begin{abstract}

We consider the minimal version of an extension of the standard
electroweak model  based on the $SU(3)_c \times SU(3)_L \times U(1)_X$
gauge symmetry (the 3-3-1 model).  We analyze the most general potential
constructed from three scalars in the triplet representation of
$SU(3)_L$, whose neutral components develop nonzero vacuum expectation
values, giving mass for all the model's massive particles. {}For different
choices of parameters, we obtain the particle spectrum for
the two symmetry breaking  scales: one where the $SU(3)_L \times
U(1)_X$ group is broken down to  $SU(2)_L\times U(1)_Y$ and a lower
scale similar to the standard model one.  Within the considerations
used, we show that the model encodes two first-order phase
transitions, respecting the pattern of  symmetry restoration.
The last transition, corresponding to the
standard electroweak one, is found to be very weak first-order, most likely 
turning second-order or a crossover in practice. However, the first transition
in this model can be strongly first-order, which might happen at a
temperature not too high above the second one. We
determine the respective critical temperatures for symmetry
restoration for the model. 

\end{abstract}

\section{Introduction}
\label{intro}

Extensive work has been dedicated to the study of the electroweak
phase transition in the standard model (SM) as well as in many of its
extensions.  This interest is based for a large part on the
possibility that it might explain the baryon asymmetry in the universe
and that this asymmetry could be produced at around the scale of the
electroweak  symmetry breaking  in the primordial hot Big Bang
universe (for reviews see, e.g.,  Ref.~\cite{reviews}).
One of the necessary conditions for a model to explain the baryon
asymmetry of the universe  is the presence of nonequilibrium
effects. In a phase transition, this can be achieved if the transition
is first-order and its strength is strong enough, in what is usually
called a strong first-order phase transition. This condition is
parameterized  by the ratio $R=\langle \phi
\rangle(T_c)/T_c$, where $\langle \phi \rangle(T_c)$ is the value for the
degenerate vacuum for the Higgs field at the critical temperature
$T_c$. A strong first-order phase transition 
is usually characterized by the condition $R>1$. In the SM this condition
cannot be achieved.  Lattice Monte Carlo numerical simulations of the
electroweak standard model~\cite{kajantie,endpoint} have shown that
there is an endpoint in the phase diagram of the model for a Higgs
mass $m_H \sim 80$ GeV, where the   phase transition is weak
first-order as the endpoint is approached  from the left and the
transition becomes a smooth crossover for larger Higgs
masses. According to recent results from the Large Hadron Collider
(LHC), from the current combined results from ATLAS and CMS
experiments~\cite{{Aad:2015zhl}}  have indicated the existence of a
Higgs boson with a mass $125.1 \pm 0.3$ GeV.  Thus, this gives no hope
of achieving the necessary conditions for producing a baryon asymmetry
in the context of the SM, since no significant departure from thermal
equilibrium can be obtained during the phase transition dynamics.
This is one of the motivations for looking for extensions of the SM
and/or alternative models and the  searches for new scalar particles
at the LHC, aiming to  reveal the ingredients needed for the strong
first-order electroweak phase transition (EWPT), as required to
produce the resulting observed baryon asymmetry. 

On the theoretical side, some  extensions of the SM have been analyzed
and the kind of  scalar was selected so as to  remedy the SM
shortcomings. These extensions used to enhance the SM are  usually
constructed with a scalar gauge singlet~\cite{MCD}, a complex scalar or a scalar
from   supersymmetric degrees of freedom (in the context of
supersymmetry extensions of the SM)~\cite{Espinosa:1993yi}. On the other hand,
there are alternative models, with a larger particle spectrum than the
SM, that predict the existence of new gauge bosons and exotic quarks
that   acquire mass from their couplings to new scalar fields.  In
particular, in this paper,  we are exploring the phenomenological
aspects  of an alternative to the SM based on the $SU(3)_c \times
SU(3)_L \times U(1)_X$ gauge symmetry, commonly known as the 3-3-1
model~\cite{PIV,FRA}. In this model, the scalars are   accommodated in
a convenient fundamental representation of the $SU(3)_L$ symmetry
group.   {}From the  electric charge operator one can select its model
version. One particular version predicts the existence of  new very
massive gauge bosons and exotic quarks.  In this work, we want to
study and better understand the possible phase transition  sequences
associated with the symmetry  breaking pattern $SU(3)_L \otimes U(1)_X
\to SU(2)_L \otimes U(1)_Y \to U(1)_{EM}$ in the 3-3-1 model and
whether it can produce the necessary conditions required for
generating a baryon asymmetry.

Let us outline some features of the model.  Although at low energies
the model has the same spectrum as the SM, it offers an explanation
for basic open questions to the SM.   In this model, the family
replication problem is solved when considering that all three families are
required for the anomaly cancellation procedure, resulting in the
number of fermion families to have to be a multiple of the  quark color
number. Considering that the QCD asymptotic freedom condition  is
valid only if the number of families of quarks is  less than five, one
concludes  that there are three generations.  Another interesting
feature of the minimal version of the model is the prediction of an
upper bound for the Weinberg angle, which follows  from a peculiar
relation between new gauge boson masses. 

The remainder of this paper is organized as follows. In
Sec.~\ref{331model},  we present the main ingredients of the 3-3-1
model related to  the spontaneous symmetry breaking (SSB).  We give a
brief exposition of its gauge, scalar and fermionic sectors, with a
description of its spectrum of particles and its main motivations for
us seeing it as an interesting and natural extension of the standard model. We also
introduce the more  general potential compatible with the given symmetry.
In Sec.~\ref{scalarspectrum}, we obtain the scalar spectrum after the SSB, 
explicitly showing the combination of the scalar fields transferring mass to the
massive gauge bosons. In Sec.~\ref{1loopsection}, we give the
expression for the quantum and thermal corrections at the one-loop
order to the tree-level potential for the model.  In
Sec.~\ref{PTpattern}, we analyze and characterize the structure of
symmetry breaking patterns in the model and we discuss our strategy for fixing the 
many parameters of the model so as to maximize the possibility of
finding a strong first-order phase transition. We study the 
temperature-dependent one-loop corrected potential as  a function of each
value expectation value of the background fields and we graphically
identify the temperature corresponding to  symmetry restoration.
{}From this analysis of the temperature dependence of the one-loop
corrected model  spectrum, we conclude that, in the 3-3-1 model, it
shows two scales for first-order phase transition, with the final
one, corresponding to the usual electroweak phase transition, as being
very weak first-order, or probably second-order in practice.
{}Finally, in Sec.~\ref{conclusions}, we give our concluding remarks.

\section{The minimal version of the 3-3-1 model}
\label{331model}

In this section we  recall the main characteristics   of the minimal
version  of the 3-3-1  model \cite{PIV,FRA} related to the
spontaneous symmetry breaking mechanism.  We start by the definition
of the electric charge operator,

\beq Q = T_3 + \beta \ T_8 + X I,
\eq
\noindent 
where $T_3$ and $T_8$ are two of the eight generators $T_i$
($i=1, \cdots, 8$)  satisfying the  $SU(3)$ algebra,  $I$ is the unit
matrix and $X$ denotes the $U(1)$ charge.  The minimal version of the model,
used in this work,  corresponds to the choice of the parameter $\beta = -
\sqrt 3$.

To generate masses for all gauge and exotic quark  fields  through
spontaneous symmetry breaking,  three triplets of scalars, denoted by
$\eta, \rho$ and $\chi$, respectively, are needed,

\begin{eqnarray} 
&&\eta =\la \eta^{0} \ \eta_{1}^{-} \ \eta_{2}^{+} \rf^T  ,
\nonumber \\
&&\rho =\la \rho^{+} \ \rho^{0} \ \rho^{++} \rf^T,
\nonumber\\ 
&&\chi =\la
\chi^{-} \ \chi^{--} \ \chi^{0}\rf^T.  
\end{eqnarray} 
The deviations of these
fields from their ground state configuration $v_\eta, v_\rho$ and
$v_\chi$, are denoted by
 
\begin{eqnarray} 
&&\eta^{0}= v_\eta + \xi_{\eta}+ i \zeta_{\eta}, 
\nonumber \\  
&&\rho^{0}=
v_\rho + \xi_{\rho} +i \zeta_{\rho},
\nonumber \\
&&\chi^{0}= v_{\chi} + \xi_{\chi}+  i \zeta_{\chi},
\end{eqnarray} 
where $\xi_{\eta,  \rho, \chi}$ and $\zeta_{\eta, \rho, \chi}$ are the deviations for
the real and imaginary components of the fields, respectively,  and we
assume that the neutral part  of each scalar develops a nonzero real
vacuum expectation value (VEV):  $\langle v_{\eta} \rangle = v_{\eta_0}$,
$\langle  v_{\rho} \rangle = v_{\rho_0}$ and $ \langle v_{\chi} \rangle = v_{\chi_0}$.   
We impose
the consistency of the model with the SM phenomenology  by adopting
$v_{\chi_0} \gg v_{\rho_0}, v_{\eta_0}$ and $v_{\rho_0}^2 +
v_{\eta_0}^2 = v_W^2= \left( 246 \, {\rm GeV}\right)^2$, where $v_{\chi_0}$
gives the  energy scale for the symmetry breaking $SU(3)_L \otimes
U(1)_X \to SU(2)_L \otimes U(1)_Y$, which is usually
assumed to be at the TeV scale, for consistency with the current
observations~\cite{CAM}. 
 
The gauge bosons, associated with the gauge symmetry $SU(3)_L$ of the model,
consist of an octet $W^i_\mu$ ($i = 1,\cdots, 8$) and a singlet $B_\mu$,
 associated with $U(1)_X$.  The model
also predicts five vector bileptons: a single charged ($V_\mu^\pm$),
a doubly charged ($Y_\mu^{\pm\pm}$)  and a new neutral gauge boson
($Z_\mu^\prime$), in addition to the charged standard model gauge
bosons ($W_\mu^\pm$), the neutral ($Z_\mu$)  and the photon
($A_\mu$). These gauge bosons are defined as

\beq 
W_\mu^\pm &=& \frac {1}{\sqrt 2}\la W_\mu^1 \mp i W_\mu^2\rf,
\\ 
V_\mu^\pm &=& \frac {1}{\sqrt 2} \la W_\mu^4 \mp i W_\mu^5 \rf,
\\
Y_\mu^{\pm\pm} &=& \frac {1}{\sqrt 2} \la W_\mu^6 \mp i W_\mu^7\rf,
\\ 
A_\mu &=& \frac{1}{\sqrt{g^2+4 g^{{\prime}^2}}} \left[ g B_\mu  +
  g^\prime\la W_\mu^3 + \sqrt 3 W_\mu^8 \rf \right], 
\\ 
Z_\mu &=& \frac{1}{\sqrt{g^2+4 g^{{\prime}^2}}}\frac{1}{\sqrt{g^2+3
    g^{{\prime}^2}}} 
\nonumber \\
&\times& \left[ g g^\prime \ B_\mu + \sqrt 3
  g^{{\prime}^2} W^8_\mu -\la g^2  + 3 g^{{\prime}^2}\rf
  W^3_\mu\right],
\\ 
Z^\prime_\mu &=& \frac{1}{\sqrt{g^2+3
    g^{{\prime}^2}}} \left[ g W^8_\mu + \sqrt 3 g^\prime B_\mu
  \right], 
\eq 
where $g$ and $g^\prime$ are the couplings defined
in the covariant derivative of the scalar fields $\Phi = \eta, \rho,
\chi,$ \beq D_\mu \Phi = \partial_\mu \Phi +i  g  W_\mu^i   T_i  \Phi
-i g^\prime B_\mu \Phi. \eq 
The new  gauge fields acquire mass at a
high scale when the  $SU(3)_L \times U(1)_X$ group  breaks down to
$SU(2)_L \times U(1)_Y$, by the $\chi$  $SU(3)_L$ scalar triplet,
while the ordinary quarks and SM gauge bosons get acceptable masses at
the next stage of SSB provided by the $\eta$ and $\rho$
triplets~\cite{TON}.  The resulting gauge boson tree-level
field-dependent masses are given by

\beq 
&& M_W = \frac g 2  \sqrt{v_{{\eta_0}}^2 + v_{{\rho_0}}^2},
\\
&& M_V =  \frac g 2   \sqrt{v_{{\eta_0}}^2 + v_{{\chi_0}}^2},
\\ 
&& M_Y =\frac g 2  \sqrt{v_{{\rho_0}}^2+ v_{{\chi_0}}^2},
\\ 
&& M_Z = \frac{g}{2 c_W}  \sqrt{v_{{\eta_0}}^2 + v_{{\rho_0}}^2},
\\
&& M_{Z^\prime} = \frac{v_{\chi_0}}{\sqrt{3}}\  \sqrt{g^ 2 + 3
        g^{\prime 2}},
\label{massZprime}
\eq
where $M_W/M_Z = c_W$, with $s_W^ 2 = 1- c_W^2 = 0.223$~\cite{PDG} 
and $g^\prime$, corresponding to the $U(1)_X$ gauge coupling, given by
$g^\prime = g \, s_W/\sqrt {1 - 4 s_W^2}$.

Let us mention that,  if the leptons are to get their masses at
tree level  within the usual Higgs mechanism, their Yukawa couplings
would require a scalar ($S$) belonging to dimension six symmetric
representation of the $SU(3)_L$ group~\cite{FOO}.  We do not evaluate
the  tiny lepton masses generated by SSB because  
they give  negligible contribution to the effective
potential.  Moreover, introducing a sextet scalar $S$ with a
background neutral field developing a  VEV, say $v_{\sigma_1} = 
\langle \sigma_1 \rangle$, would modify the previous relation between the 
field vacuum expectation values with the Weinberg scale to $v_{\sigma_1}^2 +
v_{\eta_0}^2 + v_{\rho_0}^2 = v_W^2= \left( 246 \,{\rm GeV}\right)^2$, 
but keeping the adopted estimate  $v_\chi \gg v_\eta, v_\rho, v_{\sigma_1}$.

The quark content is embedded in the extended group according to the
multiplets $\displaystyle Q_{mL} = \la d_m, u_m, j_m \rf^T_L$ and
$Q_{3 L} = \la u_3, d_3, J \rf^T_L $,  where the SM quarks are
$u_{1,2,3}$ and $d_{1,2,3}$, whereas  $J$, $j_1$ and $j_2$ are  the
exotic heavy quarks needed to complete the  fundamental
representation.  We define the background-field dependence of the top
and exotic quark masses as

\begin{eqnarray}  
&& m_{top}^2\la v_\eta, v_\rho \rf = \frac{m_{top}^2\la v_W
  \rf}{v_W^2} \la v_\eta^2 +  v_\rho^2 \rf,
\\
&& m_{Q}^2\la v_\chi
\rf =  \frac{m_{Q}^2\la v_{\chi_0} \rf}{v_{\chi_0}^2}  v_\chi^2.  
\end{eqnarray}
{}Finally, the scalar masses are obtained from the most general, gauge
invariant and renormalizable potential~\cite{TON} for the scalar
fields $\eta, \rho$ and $\chi$,

\begin{eqnarray}
V\left(\eta, \rho, \chi\right) &=& 
\mu_1^2\eta^\dagger\eta +
\mu_2^2\rho^\dagger\rho + \mu_3^2\chi^\dagger\chi +
\lambda_1\left(\eta^\dagger\eta\right)^2 
\nonumber \\
& + &
\lambda_2\left(\rho^\dagger\rho\right)^2 
+
\lambda_3\left(\chi^\dagger\chi\right)^2 
\nonumber \\
& + &  
\left[\lambda_4\left(\rho^\dagger\rho\right)
  + \lambda_5\left(\chi^\dagger\chi\right)\right]  \left(\eta^\dagger\eta\right)
\nonumber \\
& + &
\lambda_6\left(\rho^\dagger\rho\right)\left(\chi^\dagger\chi\right) +
\lambda_7\left(\rho^\dagger\eta\right)\left(\eta^\dagger\rho\right)
\nonumber  \\ & + &
\lambda_8\left(\chi^\dagger\eta\right)\left(\eta^\dagger\chi\right) +
\lambda_9\left(\rho^\dagger\chi\right)\left(\chi^\dagger\rho\right) 
\nonumber \\
& + &
\frac{1}{2} \la f_1\epsilon^{ijk}\eta_i\rho_j\chi_k +
     {\mbox{H. c.}}\rf.
\label{331pot}
\end{eqnarray}
The tree-level potential, expressed in terms of the background fields
$v_\eta$, $v_\rho$ and $v_\chi$, is 

\begin{eqnarray}
V_{\rm tree}(v_\eta,v_\rho,v_\chi) & = & \mu_1^2 v_\eta^{2}+ \mu_2^2
v_\rho^{2} +  \mu_3^2 v_\chi^{2}+ \lambda_{1} v_\eta^{4} 
\nonumber \\
&+& \lambda_2
v_\rho^{4}+ \lambda_3 v_\chi^{4}  +  \left(  \lambda_4
v_\rho^{2} +  \lambda_5 v_\chi^{2} \right) v_\eta^{2} 
\nonumber \\
&+& \lambda_6
v_\rho^{2} v_\chi^{2} +  f_1 v_\eta v_\rho v_\chi.
\label{Vtree}
\end{eqnarray}
By following a similar choice of parameters
as used, e.g., in Refs.~\cite{TON,ELM},  we fix the trilinear
coupling $ f_1$ as  $ f_1 = - {\bar f}_1 v_{\chi_0}$, where ${\bar f}_1$ is a
dimensionless constant. In particular, a common choice in the literature~\cite{TON} is
${\bar f}_1=1$. The mass parameters
$\mu_{1,2,3}$ determined by minimizing the tree-level potential in the vacuum,
which gives

\beq \mu_1^2 &=& \frac{{\bar f}_1}{2} \frac {v_{{\chi_0}}^2
  v_{{\rho_0}}}{ v_{{\eta_0}}} - 2 \lambda_{{1}} v_{{\eta_0}}^
     {2}-\lambda_{{4}} v_{{\rho_0}}^{2}-\lambda_{{5}}
     v_{{\chi_0}}^{2},
\label{mu1}\\ 
\mu_2^2 &=& \frac{{\bar f}_1}{2}  \frac {v_{{\chi_0}}^{2} v_{{\eta_0}}}
   {v_{{\rho_0}}} - 2 \lambda_{{2}} v_{{\rho_0}}^{2} -\lambda_{{4}}
   v_{{\eta_0}}^{2}-\lambda_{{6}} v_{{\chi_0}}^{2},
\label{mu2} \\ 
\mu_3^2 &=& \frac{{\bar f}_1}{2}  v_{\eta_0} v_{\rho_0} -2\lambda_{{3}}
v_{{\chi_0}}^{2}-\lambda_{{5}} v_{{\eta_0} }^{2}-\lambda_{{6}}
v_{{\rho_0}}^{2}.
\label{mu3}
\eq 

The potential (\ref{331pot})
has a too large number of, in principle, free parameters, represented
by the different possible magnitudes for the ten couplings, $\lambda_i,\,f_1$, $i=1,\ldots,9$,
the vacuum expectation values for the triplet scalars,
$v_{\chi_0}$, $v_{\eta_0}$ and $v_{\rho_0}$. Note that the constraint
$v_{\eta_0}^2 + v_{\rho_0}^2 = v_W^2$ only tells us that these two VEVs are
related to the same scale (the Weinberg scale), but does not fix the proportionality
factor between them, i.e., we can parameterize $v_{\eta_0}$ and $v_{\rho_0}$
as $v_{\eta_0} = \phi_0 \sin (\beta)$ and $v_{\rho_0} = \phi_0 \cos (\beta)$,
where $\phi_0 = v_W$, but, in principle, with an arbitrary projection angle $\beta$.
Note that a natural choice is having $v_{\eta_0} = v_{\rho_0}$, i.e.,  $\beta=45^{\circ}$,
however, in the literature there are some motivations for having  $v_{\eta_0} \neq v_{\rho_0}$,
see, e.g., Refs.~\cite{Machado:2013jca,Okada:2016whh}.
Thus, we have a total of 12 free parameters for the scalar sector,
composed of the ten couplings, the high energy scale $v_{\chi_0}$ 
associated with the first symmetry breaking  
$SU(3)_L \times  U(1)_X \to SU(2)_L\times U(1)_Y$ and the projection angle $\beta$.
In the scalar sector we can still fix one of these parameters in terms of the
others by making use of the Higgs mass $m_H$. The stability of the potential 
only constrains the possible values for the couplings. In particular, $\lambda_{1,2,3}$ 
should be positive for overall stability of the potential in the $\eta,\,\rho$ and $\chi$
directions, while the mixed couplings $\lambda_{4,5,6,7,8}$ can in principle be
negative.
 
We observe that the symmetry breaking scale for the electroweak
theory down to $U(1)_{\rm EM}$ is  governed by $v_W$. In addition, the lack of
information as regards the individual roles of $v_\eta$ and $v_\rho$ fields
in  the SSB leads us to adopt, as already mentioned above, the
polar parameterization $v_{\eta} = \phi \sin (\beta)$ and $v_{\rho} = \phi \cos (\beta)$,
such that in the vacuum, $\phi_0 = v_W$. 
Thus, after the second spontaneously symmetry breaking, where 
$SU(2)_L\times U(1)_Y \to U(1)_{EM}$, it produces  VEVs simultaneously
for both $\eta$ and $\rho$, but with an in principle arbitrary projection
angle $\beta$. 

\section {Mass spectrum for the scalars}
\label{scalarspectrum}

The scalar sector for the 3-3-1 model can be divided in  {\it CP-}even
and {\it CP-}odd scalar sectors.  The {\it CP-}even and {\it CP-}odd
scalar sectors are further composed of a neutral scalar mass matrix,
two single charged scalar and one double charged scalar matrices.
{}For  the
{\it CP-}even scalar sector, the  neutral scalar mass matrix is

\begin{widetext}
\begin{equation}  
M_{neutral}  =
\left[ \begin {array}{ccc} 
\mu_1^2+ 6 \lambda_1 v_\eta^2 +  \lambda_4 v_\rho^2 +\lambda_5 v_\chi^{2} &
2 \lambda_4  v_\eta v_\rho - \frac{{\bar f}_1}{2}  v_{\chi_0} v_\chi  &
2 \lambda_5 v_\eta\, v_\chi - \frac{{\bar f}_1}{2} v_{\chi_0} v_\rho 
\\ \noalign{\medskip}
2 \lambda_4  v_\eta v_\rho - \frac{{\bar f}_1}{2}  v_{\chi_0} v_\chi  &
\mu_2^2 + 6 \lambda_2\,{v_\rho}^2 + \lambda_4 v_\eta^2 +\lambda_6 v_\chi^{2} &
2 \lambda_6  v_\rho v_\chi - \frac{{\bar f}_1}{2}  v_{\chi_0} v_\eta 
\\ \noalign{\medskip}
2 \lambda_5 v_\eta\, v_\chi - \frac{{\bar f}_1}{2} v_{\chi_0} v_\rho  &
2 \lambda_6  v_\rho v_\chi - \frac{{\bar f}_1}{2}  v_{\chi_0} v_\eta   &
\mu_3^2 + 6 \lambda_3  v_\chi^{2} +  \lambda_5 v_\eta^2 + \lambda_6 v_\rho^2
\end {array} \right] .
\label{CPeven}
\end{equation} 
In this sector one
identifies  two single charged scalars mass matrices,  $M_{char_1}$ and
$M_{char_2}$, given, respectively, by 

\begin{equation}
 M_{char_1}=\left[ \begin {array}{cc} 
\mu_1^{2} + 2  \lambda_1 v_\eta^2+  (\lambda_4+\lambda_7) v_\rho^2 + \lambda_5  v_\chi^2 &
\lambda_7 v_\rho v_\eta + \frac{{\bar f}_1}{2} v_{\chi_0} v_\chi 
\\ \noalign{\medskip}
\lambda_7 v_\rho v_\eta + \frac{{\bar f}_1}{2} v_{\chi_0} v_\chi &
\mu_2^{2} +  2 \lambda_2  v_\rho^2+  (\lambda_4 +\lambda_7) v_\eta^2 + \lambda_6 v_\chi^2
\end{array} \right],
\end{equation} 
and 

\begin{equation}
  M_{char_2}=\left[ \begin {array}{cc} 
\mu_1^{2} + 2  \lambda_1 v_\eta^2+ \lambda_4 v_\rho^2 +  (\lambda_5+\lambda_8) v_\chi^2 &
\lambda_8 v_\eta v_\chi + \frac{{\bar f}_1}{2} v_{\chi_0} v_\rho 
\\ \noalign{\medskip}
\lambda_8 v_\eta v_\chi + \frac{{\bar f}_1}{2} v_{\chi_0} v_\rho  &
\mu_3^{2} +  2 \lambda_3  v_\chi^2  + \lambda_6 v_\rho^2 +  (\lambda_5 +\lambda_8) v_\eta^2
\end{array} \right].
\end{equation}  
There is a double charge $M_{double}$ scalar mass matrix, given by

\begin{equation}
M_{double}= \left[ \begin {array}{cc} 
\mu_2^{2} + 2 \lambda_2 v_\rho^2 + \lambda_4 v_\eta^2 + \left( \lambda_6 + \lambda_9 \right) v_\chi^2 & 
\lambda_9 v_\rho v_\chi + \frac{{\bar f}_1}{2} v_{\chi_0} v_\eta
\\ \noalign{\medskip}
\lambda_9 v_\rho v_\chi + \frac{{\bar f}_1}{2} v_{\chi_0} v_\eta &
\mu_3^{2} +2\lambda_3 v_\chi^{2} + \lambda_5 v_\eta^2 + \left(\lambda_6 + \lambda_9\right) v_\rho^2
\end{array} 
\right].
\end{equation}  
Next, we give the {\it CP-}odd scalar sector. The mass matrix of neutral scalars is given by

\begin{equation}
 M^{CP}_{neutral}=\left[ \begin {array}{ccc} 
\mu_1^{2} + 2 \lambda_1 \,v_\eta^{2} +\lambda_4 v_\rho^2 + \lambda_5 v_\chi^{2} &
\frac{{\bar f}_1}{2}   v_{\chi_0} v_\chi &
\frac{{\bar f}_1}{2}   v_{\chi_0} v_\rho
\\ \noalign{\medskip}
\frac{{\bar f}_1}{2}   v_{\chi_0} v_\chi &
\mu_2^{2}+ 2 \lambda_2 v_\rho^2+  \lambda_4 v_\eta^2 +   \lambda_6 v_\chi^2 &
\frac{{\bar f}_1}{2}     v_{\chi_0}   v_\eta 
\\ \noalign{\medskip}
\frac{{\bar f}_1}{2}   v_{\chi_0} v_\rho& 
\frac{{\bar f}_1}{2}     v_{\chi_0} v_\eta &
\mu_3^{2} + 2 \lambda_3 v_\chi^2 +  \lambda_5 v_\eta^2 + \lambda_6 v_\rho^2
\end{array}\right]. 
\end{equation} 
In this sector there are two mass matrices
of single charged scalars, given, respectively, by

\begin{equation}
M^{CP}_{char_1}= \left[ \begin {array}{cc} 
\mu_1^{2}+ 2 \lambda_1 v_\eta^2 + \left(\lambda_4+\lambda_7 \right) v_\rho^2+ \lambda_5 v_\chi^2 &
-\lambda_7 v_\eta v_\rho - \frac{{\bar f}_1}{2} v_{\chi_0} v_\chi
\\ \noalign{\medskip}
-\lambda_7 v_\eta v_\rho - \frac{{\bar f}_1}{2} v_{\chi_0} v_\chi & 
\mu_2^{2} +  2 \lambda_2 v_\rho^2  + \left( \lambda_4+ \lambda_7 \right) v_\eta^2 + \lambda_6 v_\chi^2
\end{array} \right], 
\end{equation} 
and

\begin{equation}
M^{CP}_{char_2}= \left[ \begin {array}{cc} 
\mu_1^{2}+ 2 \lambda_1 v_\eta^2 + \lambda_4 v_\rho^2 + \left(\lambda_5+\lambda_8 \right) v_\chi^2 &
-\lambda_8 v_\eta v_\chi - \frac{{\bar f}_1}{2} v_{\chi_0} v_\rho
\\ \noalign{\medskip}
-\lambda_8 v_\eta v_\chi - \frac{{\bar f}_1}{2} v_{\chi_0} v_\rho & 
\mu_3^{2} +  2 \lambda_3 v_\chi^2  + \lambda_6 v_\rho^2 + \left( \lambda_5+ \lambda_8 \right) v_\eta^2 
\end {array} \right], 
\end{equation}  
and a matrix for doubly
charged {\it CP-}odd scalars,

\begin{equation}
M^{CP}_{double}= \left[ \begin {array}{cc}
\mu_2^{2}+ 2 \lambda_2 v_\rho^2 + \lambda_4 v_\eta^2 + \left(\lambda_6+\lambda_9 \right) v_\chi^2 &
-\lambda_9 v_\rho v_\chi - \frac{{\bar f}_1}{2} v_{\chi_0} v_\eta
\\ \noalign{\medskip}
-\lambda_9 v_\rho v_\chi - \frac{{\bar f}_1}{2} v_{\chi_0} v_\eta & 
\mu_3^{2} +  2 \lambda_3 v_\chi^2  + \lambda_5 v_\eta^2 + \left( \lambda_6+ \lambda_9 \right) v_\rho^2 
\end{array} \right]. 
\end{equation} 
\end{widetext}

The scalar and gauge boson
masses depend on two VEVs ($v_W$ and $v_{\chi_0}$), on the projection angle $\beta$ 
and on the ten couplings. We obtain the scalar  masses by diagonalizing
the corresponding $3\times3$ and $2 \times 2$ mass matrices given
above.

We have in mind previous analyses~\cite{TON,ELM,JOS} predicting  that
the neutral scalar {\it CP}-even sector  must contain a low mass
component corresponding to the SM Higgs particle $H$.  In addition, in
this model, one expects two neutral scalars, $H_1^0$  and $H_2^0$, in
the {\it CP-}even sector. In the {\it CP-}odd sector, there is another 
neutral scalar, $H_{cp}^0$, along with two Goldstone bosons.  
{}For the charged states, one expects six  massive scalars, $H_1^\pm, H_2^\pm$ and $H^{\pm \pm}$
and another six Goldstone bosons. All the eight Goldstone bosons give mass to the massive gauge
bosons, i.e., the SM gauge bosons $Z$ and $W^{\pm}$ and the additional heavy
bosons predicted by the present version of the model, $Z^\prime$, $V^\pm$ and $Y^{\pm\pm}$.
The four singly charged  massive gauge fields
($W^\pm$ and $V^\pm$), two doubly charged  massive gauge fields
($Y^{\pm \pm}$) and two neutral massive gauge fields ($Z$ and
$Z^\prime$).  These gauge fields have to obtain mass from the Higgs
mechanism occurring at the electroweak $v_W$ scale and at the
$v_{\chi_0}$ higher energy scale. 

Our aim is to obtain the scalars, exotic quarks, and gauge boson masses
using minimum arbitrariness.  In order to fit  the
Higgs mass and the Goldstone fields with some  set of parameters, one
must respect the recent gauge boson $Z^\prime$ mass  lower limit
determined  from the upper limit on the ATLAS/LHC electron and muon
production cross-section~\cite{CAM} (note also that there are also
similar constrains for the $Z'$ from calculations of the muon magnetic 
moment~\cite{Kelso:2014qka}).
By diagonalizing the
$M_{neutral}$ matrix  it is possible to respect the LHC constraint~\cite{CAM},
$Z_{331 minimal}  >  2.93$ TeV and to reproduce the SM Higgs
mass. This in turn, from the expression for $Z'$, Eq.~(\ref{massZprime}),
leads to a lower bound on the scale, $v_{\chi_0} \gtrsim 3$ TeV.

Once a given set of couplings are given, we obtain the  whole scalar spectrum from
the eigenvalues of the corresponding   matrices calculated in the vacuum,
$v_\eta= v_{\eta_0} \equiv v_W \sin(\beta)$, $v_\rho= v_{\rho_0} \equiv v_W \cos(\beta)$
and $v_\chi=v_{\chi_0}$. It
results in nine scalars constructed from the real components of the fields.
The SM Higgs is constructed from the combination
$\eta^0 - \rho^0$. The other
eight heavy scalars, namely  $H_1^0$ and  $H_2^0$, are related
to the
$\eta^0 - \chi^0$ and $\rho^0 - \chi^0$ combinations, respectively,
$H_1^\pm$ and $H_2^\pm$ are related  to the combinations
$\eta_1^\pm-\chi^\pm$ and $\eta_2^\pm-\chi^\pm$, respectively, and $H^{\pm
  \pm}$, which is  related to the $\rho^{\pm \pm}-\chi^{\pm \pm}$
combination.  In the {\it CP-}odd sector, there is one heavy neutral
$H_{cp}^0$, which is related to the imaginary part of the neutral field components.
The gauge bosons $W^\pm, V^\pm$ and $Y^{\pm \pm}$ acquire their masses from
the imaginary part of the fields in combinations $\eta_1^\pm-\rho^\pm$,
$\eta_2^\pm-\chi^\pm$ and $\rho^{\pm \pm} - \chi^{\pm \pm}$, respectively.
The neutral gauge bosons $Z$ and $Z'$ get their masses from
the imaginary part of the fields in the $\eta^0-\rho^0$ and
$\rho^0-\chi^0$  combinations, respectively.

As already observed, the model has a too large number of free parameters,
which makes it an almost impossible job to study the complete
parameter region allowed.
Since our objective in this work is to determine whether a strong first-order 
phase transition in the model is possible, our strategy for fixing the many couplings 
is then chosen so as to maximize this goal.
{}For this purpose, we can borrow some of the lessons already learned when studying the
phase transition in the SM and other extensions of 
it (see, e.g., Refs.~\cite{Espinosa:1992kf,Ham:2004cf,Ahriche:2007jp} and
references therein). To satisfy the usual criterion for a
strong first-order phase transition, namely that the ratio of the field expectation
value at the critical temperature and the critical temperature be
larger than one, $v(T_c)/T_c > 1$, we need, optimally, either a larger VEV at $T_c$
and/or a small $T_c$. Typically $T_c$ is constrained by the scale,  
$T_c \propto v_0$, which for us is rather large (recalling that $v_{\chi_0}
\gtrsim 3$ TeV and $v_{\eta_0}$ and $v_{\rho_0}$ are constrained by the
Weinberg scale. Since in general the VEV is given in terms of a combination
of couplings and masses, $v \propto m_i/\sqrt{\lambda_i}$, an ideally situation
is to try to work with the smallest couplings possible. There is, however, a
trade off. Too small couplings lead in general to a light particle spectrum,
which for us is still limited by the scales and observational bounds
(in particular, other scalars than the SM Higgs are expected to be
sufficiently heavy for not being detected yet). 
The Higgs mass $m_H$ itself is the only limiting observational quantity 
we have in the scalar sector. Since the Higgs is a mass eigenvalue 
for the {\it CP-}even scalar neutral matrix, Eq.~(\ref{CPeven}),
it only (weakly) constrains the couplings $\lambda_{i}$, $i=1,\ldots,6$
and ${\bar f}_1$. 

In the analysis below, we fix $m_H = 125$ GeV and work
with four different sets of choices for couplings. Other possibilities
can be shown to fall in one of these sets. In each of the sets used, we
look for the ideal conditions for having a strong first-order phase
transition, which, as mentioned above, favors the smallest 
choice of couplings in general. 
In all the sets we consider below, we found more convenient to vary the inter-couplings
between the different fields,
$\lambda_4, \, \lambda_5$ and $\lambda_6$, due to their relation to the
SM Higgs mass (the other inter-couplings $\lambda_7, \, \lambda_8$ and $\lambda_9$,
only appear on the heavy charged scalars and are unconstrained by the Higgs).
We will consider the following four large sets of parameters:

\begin{description}
\item (a) Set I: The couplings 
$\lambda_i$, with $i=1,2,3,5,6,7,8,9$, such that
$\lambda_i = \lambda$ varied together with $\lambda_4$;
\item (b) Set II: The couplings 
$\lambda_i$, with $i=1,2,3,4,6,7,8,9$, such that
$\lambda_i = \lambda$ varied together with $\lambda_5$;
\item (c) Set III: The couplings 
$\lambda_i$, with $i=1,2,3,4,5,7,8,9$, such that
$\lambda_i = \lambda$ varied together with $\lambda_6$;
\item (d) Set IV: The couplings 
$\lambda_i$, with $i=1,2,3,7,8,9$,such that
$\lambda_i = \lambda$ and  $\lambda_4 = \lambda_5=\lambda_6={\bar \lambda}$,
which are then varied. 
\end{description}
{}For all sets we have fixed ${\bar f}_1=1$. This is motivated by the fact that
${\bar f}_1$ determines the asymmetry of the potential in the $\chi$ direction
and the more asymmetrical is the potential, the more we expect to have a stronger
first-order phase transition. Note also that values of couplings larger than one
can make us enter in a nonperturbative regime of parameters. We avoid this
situation here, since we work only at the one-loop level for the effective potential
for the model (see next section). In particular, we have explicitly checked that
smaller values of ${\bar f}_1$ always lead to weaker transitions. 
In addition, for each of the sets explained above, we have chosen to work
with the higher energy scale $v_{\chi_0}$ with values $v_{\chi_0}= 3$ , 4  and
5 TeV, satisfying the current constraints on the $Z'$ mass, as already mentioned. 
Likewise, for the projection angle $\beta$, we have considered for each of the sets 
and values of the scale, the three values $\beta = 30^{\circ}, \, 45^{\circ},\, 60^{\circ}$. Again,
we have explicitly verified that larger asymmetries on the $v_{\eta}$ and
$v_\rho$ directions are either disfavored, or there is a trade off, since as 
we decrease the projection in one direction, there is a compensation by the
increase of the projection in the other direction (recalling again
that  $v_{\eta}$ and $v_\rho$ are constrained by the Weinberg scale $v_W$).
Nonetheless, our analysis shows that the symmetrical case
$v_\eta = v_\rho$, i.e., $\beta = 45^{\circ}$, tends to be favored as far as the
strength of the transition is concerned. 

\begin{center}
\begin{figure*}
\subfigure[Set I, where $\lambda_i\equiv \lambda$, with $i=1,2,3,5,6,7,8,9$.]
{\includegraphics[width=7cm,height=5cm]{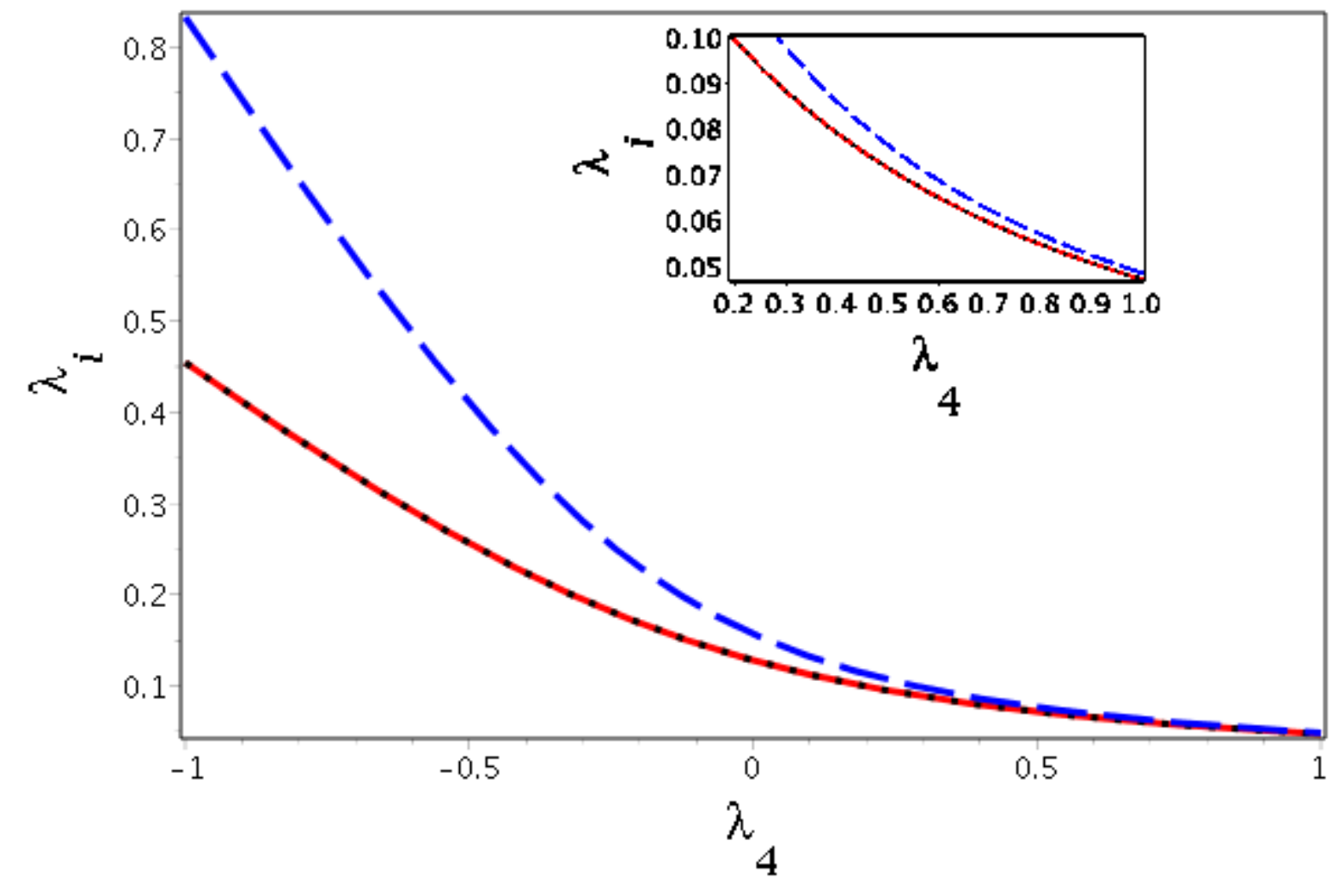}}\hspace{0.5cm}
\subfigure[Set II, where $\lambda_i\equiv \lambda$, with $i=1,2,3,4,6,7,8,9$.]
{\includegraphics[width=7cm,height=5cm]{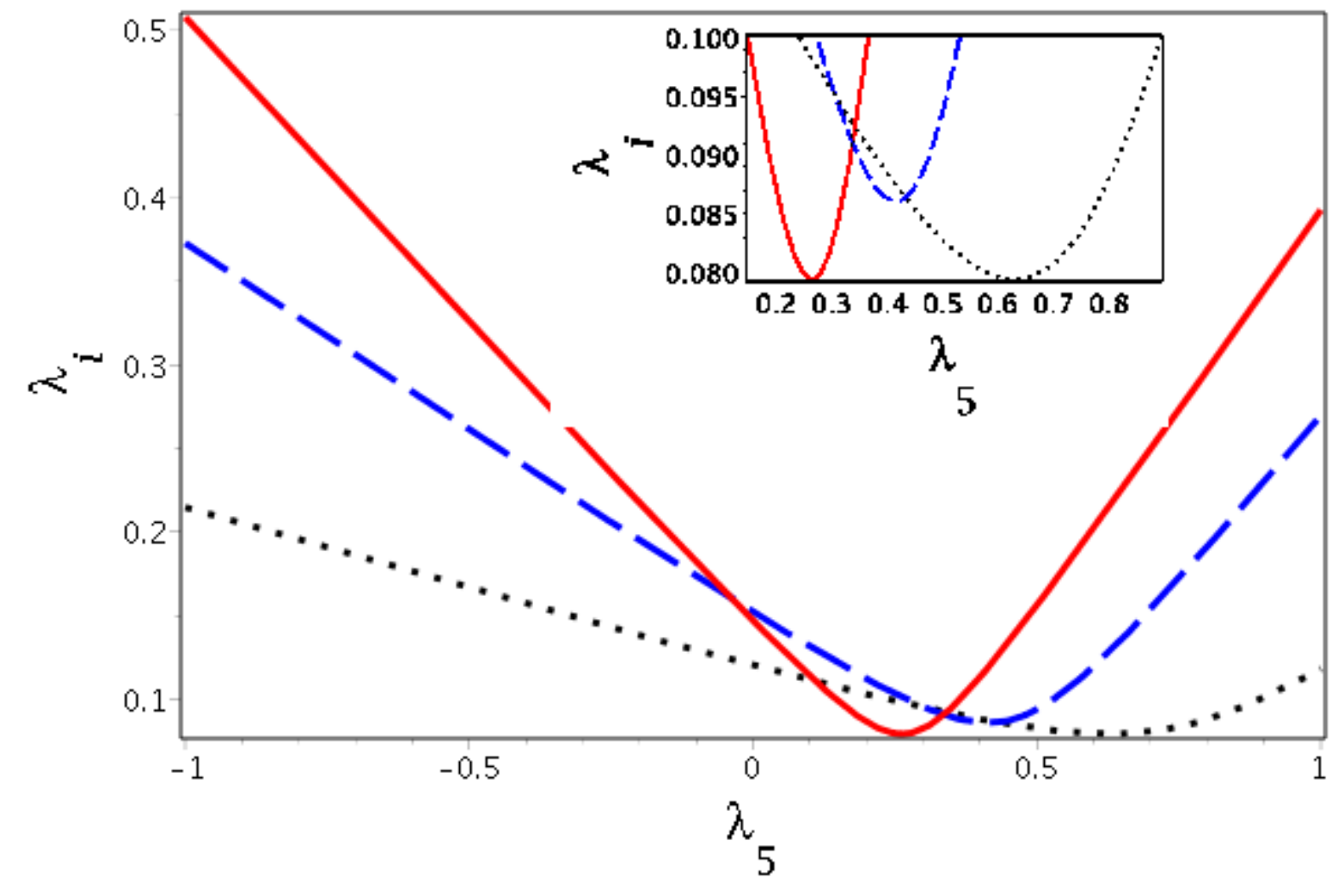}}\\
\subfigure[Set III, where $\lambda_i\equiv \lambda$, with $i=1,2,3,4,5,7,8,9$.]
{\includegraphics[width=7cm,height=5cm]{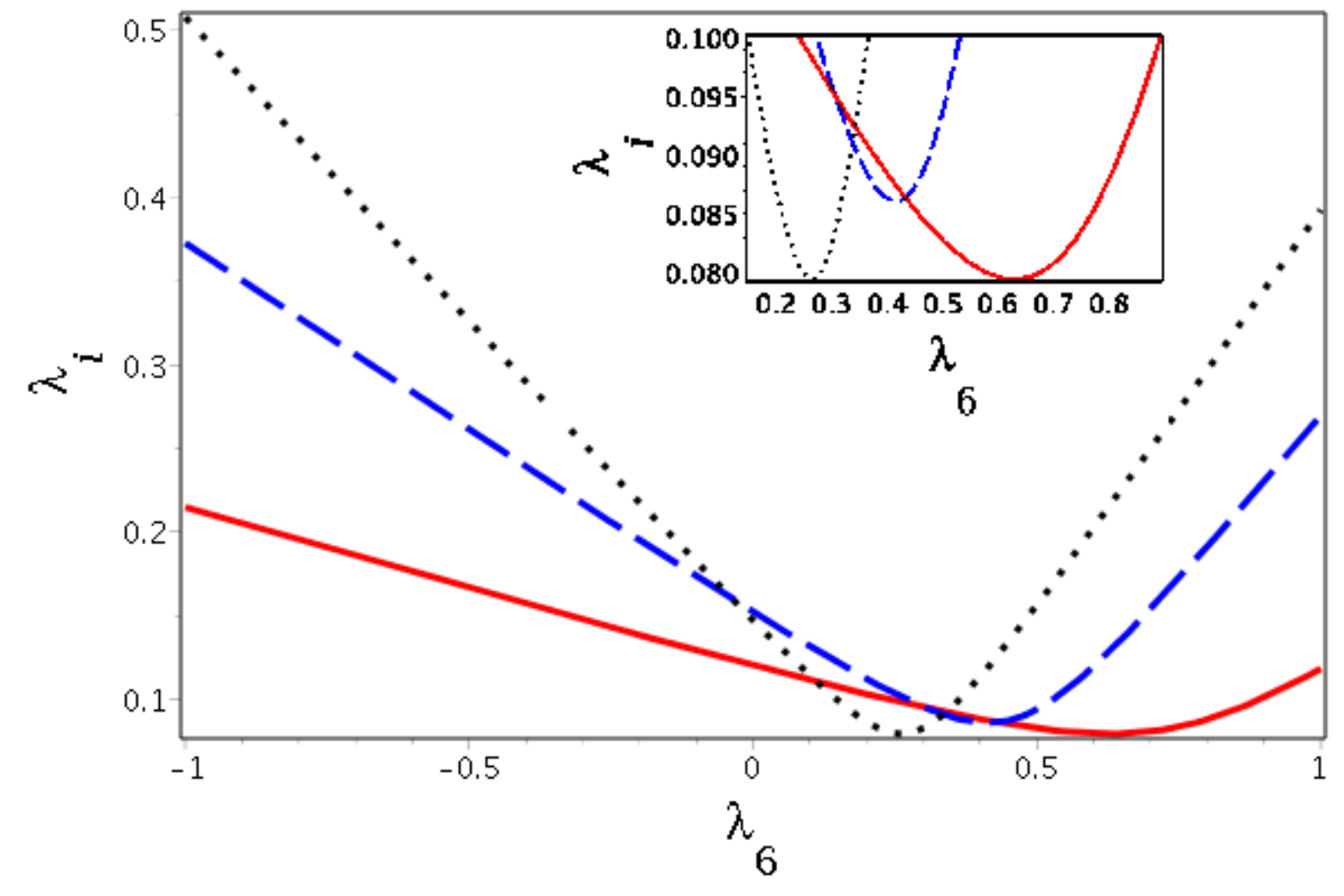}}\hspace{0.5cm}
\subfigure[Set IV, where $\lambda_i\equiv \lambda$, with $i=1,2,3,7,8,9$ and  
$\lambda_j\equiv {\bar\lambda}$, with $j=4,5,6$.]
{\includegraphics[width=7cm,height=5cm]{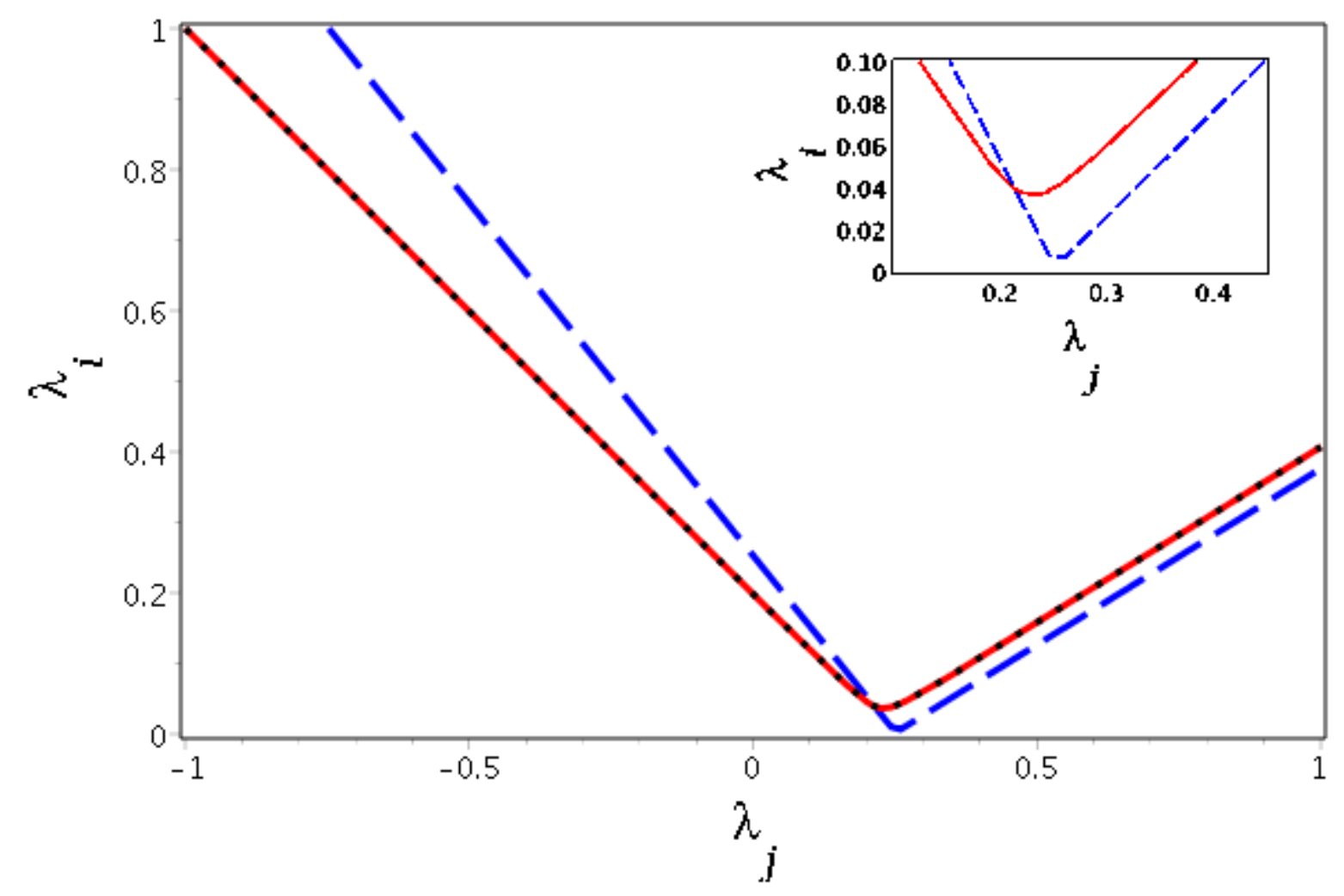}}
\caption{Set of couplings fitting the SM Higgs mass. Dotted lines are for projection 
angle $\beta = 30^{\circ}$, dashed lines is for $\beta = 45^{\circ}$ and solid lines
for $\beta = 60^{\circ}$. The insets show a region around the minimal values for the
couplings. }
\label{figsets}
\end{figure*}
\end{center}

Each one of the parameters in the above four sets is then chosen so as
to satisfy the Higgs mass  $m_H = 125$ GeV. The resulting relations
between these couplings subject to this constraint are shown in
{}Figs.~\ref{figsets}(a), \ref{figsets}(b), \ref{figsets}(c) and \ref{figsets}(d).

We note from the results shown in {}Fig.~\ref{figsets} that for sets
II, III and IV, there are always minimal values for the couplings.
In the case of the set I, $\lambda_i$ is a decreasing function
of $\lambda_4$. 

{}For illustration, the resulting scalar mass spectrum in each set is shown in the 
Table~\ref{spectrum}. {}For convenience, we show only the values at the minimal 
values of couplings in the case of the sets II, III and IV shown in {}Fig.~\ref{figsets}. 
In the case of set I, the values of the masses are for $\lambda_4=1$, which we take 
as the limit for a "perturbative" coupling.  Note that exchanging $\beta =30^{\circ}$ by 
$\beta= 60^{\circ}$ corresponds to change $v_\eta$ by $v_\rho$. The dependence of $M_2^\pm$ 
on the set  \{$v_{\rho^0}, v_{\eta^0}$\} is the same as the dependence of $M^{\pm\pm}$ 
on \{$v_{\eta^0}, v_{\rho^0}$\}. 
As a consequence, the mass of the singly charged scalar $H_2^\pm$ for $\beta = 30^{\circ}$ is 
the same as that for the double charged $H^{\pm \pm}$ for $\beta = 60^{\circ}$.
{}For completeness, the
heavy gauge boson masses $M_{Z^\prime}$, $M_{V^\pm}$  and $M_{Y^{\pm \pm}}$ are  
shown in Table~\ref{spectrumgauge}, for the cases of $v_{\chi_0}=3,\, 4$ and $5$ TeV. 
Here again the dependence of $M_{V^\pm}$ on the set \{$v_{\eta^0}, v_{\rho^0}$\} is the same 
as that of $M_{Y^{\pm\pm}}$ on the set \{$v_{\eta^0}, v_{\rho^0}$\} and so we observe 
that $M_{V^\pm}$ for $\beta=30^{\circ}$ is equal to $M_{Y^{\pm\pm}}$ for $\beta=60^{\circ}$, 
for any  $v_{\chi^0}$ value.

\begin{table}[H]
\caption{The masses (in units of TeV)  for the additional scalars in the model. The
scale $v_{\chi_0}$ as been fixed in the value of 3 TeV. }
\label{spectrum}
\begin{center}
\begin{tabular}{c|c|c|c|c|c|c|c}
\hline 
Set  &  $\beta$ &  $M_{H_1^0}$ &  $M_{H_2^0}$ & $M_{H_cp^0}$  & $M_{H_1^\pm}$ &  $M_{H_2^\pm}$ & $M_{H^{\pm \pm}}$  
\\ \hline   
  &  $30^{\circ}$ & 3.218 & 1.320  & 3.226 & 3.224 & 2.869  &   1.743
\\  
I  &  $45^{\circ}$ &  2.991 & 1.345  & 3.003 & 3.000 & 2.225 &  2.225 
\\   
  &  $60^{\circ}$ & 3.218 & 1.320 & 3.226  & 3.224 & 1.743 & 2.869    
\\ \hline   
  &  $30^{\circ}$ & 3.225 & 1.694  & 3.226 & 3.224 & 2.920 &   1.825
\\  
 II  &  $45^{\circ}$ &  3.003 & 1.762  & 3.003 & 3.001 & 2.301 & 2.301  
\\   
  &  $60^{\circ}$ & 3.226 & 1.692 & 3.226  & 3.224 & 1.825 & 2.920    
\\ \hline   
  &  $30^{\circ}$ & 3.226  & 1.692  & 3.226 & 3.224  & 2.920 &  1.825 
\\  
 III  &  $45^{\circ}$ & 3.003 & 1.762  & 3.003 & 3.001 & 2.301 &  2.301 
\\   
  &  $60^{\circ}$ & 3.225 & 1.694 & 3.226  & 3.224 & 1.825 &  2.920   
\\ \hline   
  &  $30^{\circ}$ & 3.223 & 1.148  & 3.226 & 3.224 & 2.852 &  1.715 
\\  
 IV  &  $45^{\circ}$ & 2.998 & 0.383  & 3.003 & 3.000 & 2.133 & 2.133  
\\   
  &  $60^{\circ}$ & 3.223 & 1.148 & 3.226  & 3.224 & 1.715 & 2.852    
\\ \hline   
\end{tabular}
\end{center}
\end{table}

\begin{table}[H]
\caption{The masses (in units of TeV)  for the
heavy gauge bosons $Z^\prime$, $V^\pm$  and $Y^{\pm \pm}$. }
\label{spectrumgauge}
\begin{center}
\begin{tabular}{c|c|c|c|c}
\hline 
$v_{\chi_0}$ (TeV)  &  $\beta$ &  $M_Z'$ &  $M_{V^\pm}$ &  $M_{Y^{\pm \pm}}$  
\\ \hline   
  &  $30^{\circ}$ &   & 0.981  & 0.983 
\\  
3  &  $45^{\circ}$ &  3.035 & 0.982  & 0.982
\\   
  &  $60^{\circ}$ &  & 0.983 & 0.981  
\\ \hline   
  &  $30^{\circ}$ &  & 1.307  & 1.309 
\\  
4  &  $45^{\circ}$ & 4.047 & 1.308  & 1.308   
\\   
  &  $60^{\circ}$ &  & 1.309 & 1.307    
\\ \hline   
  &  $30^{\circ}$ &   & 1.634  & 1.635 
\\  
5  &  $45^{\circ}$ & 5.059 & 1.634  & 1.634 
\\   
  &  $60^{\circ}$ &  & 1.635 & 1.634    
\\ \hline   
\end{tabular}
\end{center}
\end{table}

In the next section we introduce the quantum and thermal corrections at
the one-loop level for the effective potential in the 3-3-1 model and
we analyze its temperature dependence, for the different set of parameters
explained above, obtaining the symmetry
restoration temperatures. As input, we use the SM values for the masses of the 
Higgs, quark top, gauge boson $Z$ and $W^\pm$:
$m_H=125$ GeV,  $m_{top} = 173 .21$ GeV, $M_Z = 91.19$ GeV and $M_{W^\pm} =
80.39$ GeV. The exotic heavy quarks masses
have been fixed as $m_Q \la v_{\chi_0}\rf = v_{\chi_0}/2$.

\section{The one-loop effective potential for the 3-3-1 model}
\label{1loopsection}

The effective potential is expressed as function of the
background values for the scalars $\langle \eta \rangle=v_\eta$,
$\langle \rho \rangle = v_\rho$ and  $\langle \chi \rangle =
v_\chi$. It depends on the loop contributions from the gauge bosons, 
through their tree-level background-field-dependent
masses,  as well as  those from  fermions that can give a significant
contribution to the effective potential,  namely the top quark (t) and
the three exotic heavy quarks (Q).  {}Finally, we also have to add the
contributions from the SM Higgs and from the nine scalars that  become
heavy in the vacuum after the first SSB. Besides, when choosing a
gauge other than the unitary gauge, we have also to include the
contributions from the eight Goldstone bosons.  In this work, as is
usual in the literature (see, e.g., Ref.~\cite{Espinosa:1992kf} and
references therein), we give the expression for the effective
potential in the t'Hooft-Landau gauge.

The effective potential in terms of the background fields is expressed
as

\begin{eqnarray}
V_{\rm eff}(v_\eta,v_\rho,v_\chi, T) &=& V_{\rm
  tree}(v_\eta,v_\rho,v_\chi) +\Delta V_0 (v_\eta,v_\rho,v_\chi) 
\nonumber \\
&+& \Delta V_T (v_\eta,v_\rho,v_\chi, T),
\label{Veff}
\end{eqnarray}
where $V_{\rm tree}$ is the tree-level potential, Eq.~(\ref{Vtree}),
\break
$\Delta V_0 (v_\eta,v_\rho,v_\chi)$ is the zero temperature (quantum)
contribution for the one-loop effective potential, while \break
$\Delta V_T
(v_\eta,v_\rho,v_\chi, T)$ is the finite temperature contribution at
the one-loop level.

The one-loop quantum contribution $\Delta V_0$ is ultraviolet
divergent and needs to be renormalized.  In the cutoff regularization
scheme with subtraction point chosen at the scalar vacuum expectation
values (thus preserving the values of $v_{\eta_0}, v_{\rho_0}$ and
$v_{\chi_0}$), $\Delta V_0 (v_\eta,v_\rho,v_\chi)$ is given
by~\cite{Espinosa:1992kf}

\begin{eqnarray}
\label{potV0}
\lefteqn{\Delta V_0 (v_\eta,v_\rho,v_\chi)=\frac{1}{64\pi^2}\sum_{i} n_i
\left\{m_i^4(v)\left[ \ln
  \frac{m_i^2(v)}{m_i^2(v_0)}-\frac{3}{2}\right] 
\right. }
\nonumber \\
&& + \left. 2 m_i^2(v_0) m_i^2(v)
\frac{}{}\right\} \nonumber \\ &&+\frac{1}{64\pi^2} \sum_{G} n_G \, m_G^4(v)\left[
  \ln \frac{m_G^2(v)}{m_H^2(v_0)}-\frac{3}{2}\right],
\end{eqnarray}
where $n_i$ in Eq.~(\ref{potV0}) denotes the field degrees of freedom:
The massive charged gauge bosons have $n_i=6$  (e.g., $W^{\pm},
V^\pm, Y^{\pm \pm}$), the neutral massive gauge bosons have $n_i=3$ (e.g.,
$Z,Z'$), the heavy quarks have $ n_i=-12$ (e.g.,  the top $t$ and the
three new exotic quarks $Q$) and each of the neutral and charged scalars has $n_i=1$.  
The last sum in Eq.~(\ref{potV0}) is over the Goldstone modes, each one contributing
with $n_G=1$.  {}Finally, the masses $m_{i(G)}(v)$ and
$m_{i(G)}(v_0)$, with $v \equiv v_\eta, v_\rho, v_\chi$ and $v_0
\equiv v_{\eta_0}, v_{\rho_0}, v_{\chi_0}$, stand for the particle
masses computed at the background and vacuum expectation values,
respectively.

\begin{center}
\begin{figure}[htb!]
\subfigure[]{\includegraphics[width=7cm,height=5cm]{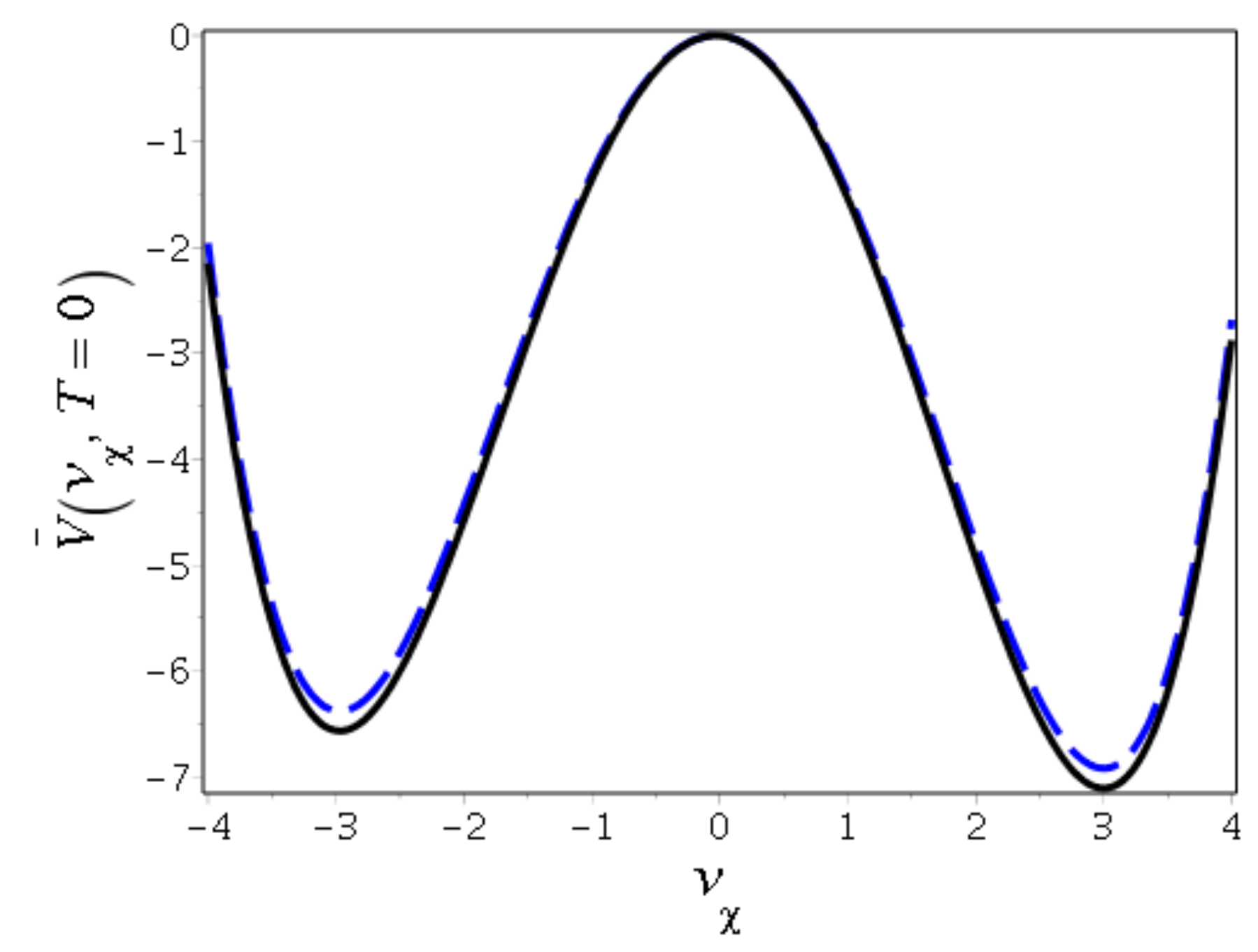}}\hspace{0.5cm}
\subfigure[]{\includegraphics[width=7cm,height=5cm]{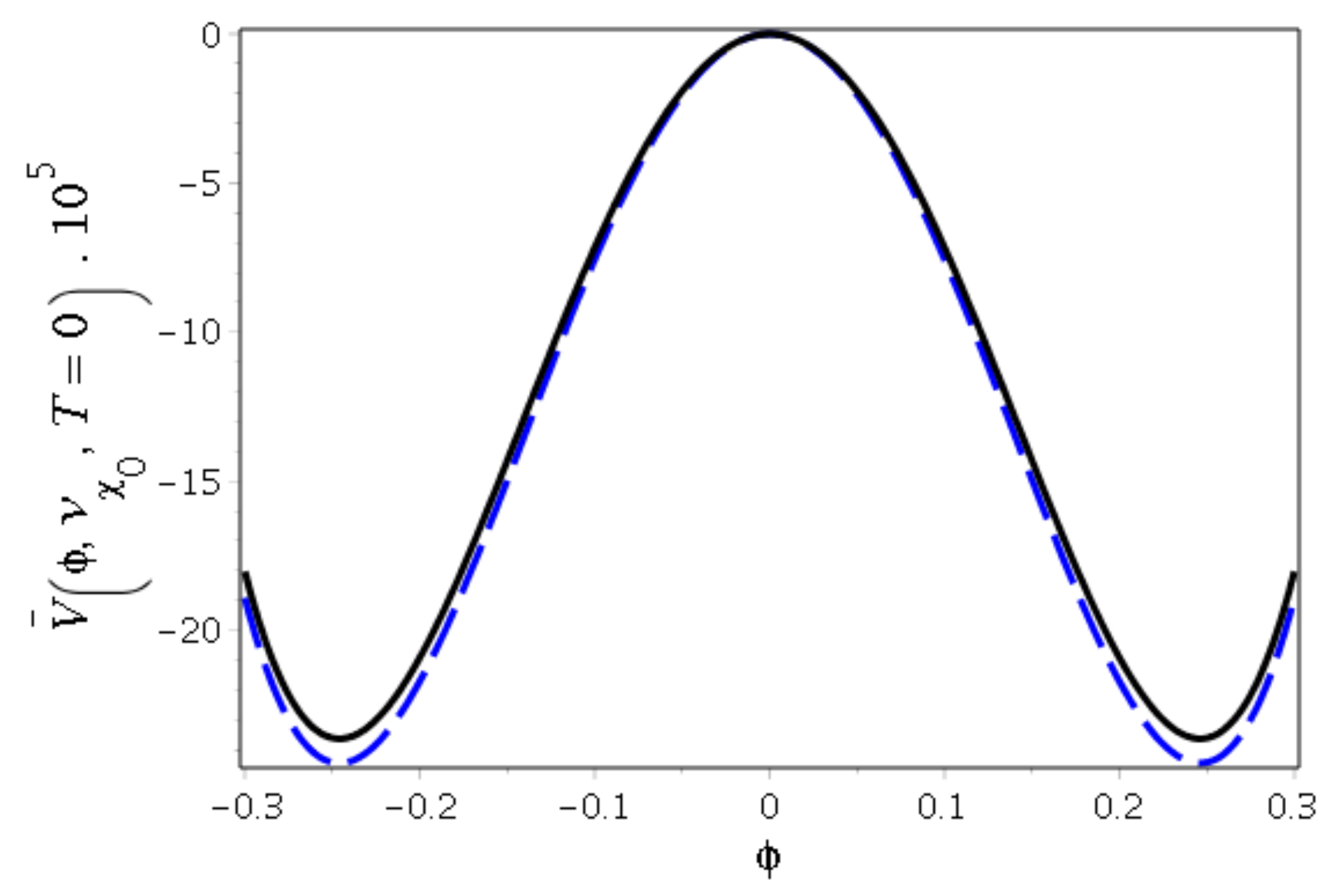}}
\caption{The vacuum subtracted tree-level potential (solid line) and
  the one-loop quantum corrected effective potential (dashed line) in
  the direction of the background scalar field $v_\chi$ (at $v_\eta =
  v_\rho=v_W/\sqrt{2}$) (panel (a)) and  in the direction of the
  background scalar field $v_\eta = v_\rho$ (panel (b)), expressed in terms
of $\phi = \pm\sqrt{v_\eta^2 + v_\rho^2}$.  The potential is in units
  of TeV$^4$ and the background fields are in units of TeV. The parameters used
are those from set II, for $\tan(\beta)=1$ and $v_{\chi_0}=3$ TeV.}
\label{figpot}
\end{figure}
\end{center}

As an example, in {}Fig.~\ref{figpot} we show both the tree-level and the one-loop
quantum corrected potentials in the case of $\tan(\beta)=1$, i.e., for $ v_\rho =v_\eta $,
in the set  II case of parameters explained in the previous section, taken at the minimal
values of couplings. We have also
considered the scale $v_{\chi_0}= 3$ TeV. {}For convenience of presentation, we have
subtracted from the potential the vacuum contribution at the origin
(corresponding to an overall shift of the whole potential).

The finite temperature contribution in Eq. (\ref{Veff}), \break 
$\Delta V_T(v_\eta,v_\rho,v_\chi, T)$,  is given by~\cite{kapusta}

\begin{eqnarray}
\label{potVT}
\Delta V_T (v_\eta,v_\rho,v_\chi,
T) &=&\frac{T^4}{2\pi^2}\left[\sum_{i={\rm bosons}} n_i
  J_B[m_i^2(v)/T^2] \right.
\nonumber \\
&+& \left.   \sum_{i={\rm fermions}} n_i J_F[m_i^2(v)/T^2]
  \right],
\end{eqnarray}
where the functions $J_B$ and $J_F$ are defined as

\begin{equation}
\label{jb}
J_B(y)=\int_0^{\infty} dx\ x^2\ln\left[1-e^{-\sqrt{x^2+y^2}}\right]\;,
\end{equation}
and

\begin{equation}
\label{jf}
J_F(y)=\int_0^{\infty} dx\ x^2\ln\left[1+e^{-\sqrt{x^2+y^2}}\right]\;.
\end{equation}

The thermal bosonic one-loop integral (\ref{jb}),  admits a
high-temperature expansion, for $y \ll 1$ (where $y=m(v)/T$). It is
given by~\cite{kapusta}

\begin{eqnarray}
\label{jbexp}
\lefteqn{ J_B(y)  =  -\frac{\pi^4}{45}+ \frac{\pi^2}{12}y^2-\frac{\pi}{6}
\left(y^2\right)^{3/2}-\frac{1}{32} y^4\ln\frac{y^2}{a_b} } \nonumber
\\ & &
-2\pi^{7/2}\sum_{\ell=1}^{\infty}(-1)^{\ell}\frac{\zeta(2\ell+1)}
{(\ell+1)!}\Gamma\left(\ell+\frac{1}{2}\right)
\left(\frac{y^2}{4\pi^2} \right)^{\ell+2} ,
\end{eqnarray}
while the thermal fermionic one-loop integral (\ref{jf}), for $y \ll
1$, can likewise be expressed as

\begin{eqnarray}
\label{jfexp}
\lefteqn{ J_F(y)  =  \frac{7\pi^4}{360}- \frac{\pi^2}{24}y^2-\frac{y^4}{32}
\ln \frac{y^2}{a_f} }
\nonumber \\ 
&&-\frac{\pi^{7/2}}{4}\sum_{\ell=1}^{\infty}(-1)^{\ell}
\frac{\zeta(2\ell+1)}{(\ell+1)!} 
 \left(1-2^{-2\ell-1}\right) 
\nonumber \\
&& \times
\Gamma\left(\ell+\frac{1}{2}\right)\left(\frac{y^2}{\pi^2}
\right)^{\ell+2}.
\end{eqnarray}
In the above expressions, $a_b=16\pi^2\exp(3/2-2\gamma_E)$  (and $\ln
a_b=5.4076$), $a_f=\pi^2\exp(3/2-2\gamma_E)$  (or $\ln a_f=2.6351$)
and $\zeta$ is the Riemann $\zeta$-function.

In the opposite regime of a low-temperature, $y> 1$, the  integrals
(\ref{jb}) and (\ref{jf}) are well approximated by

\begin{equation}
J_{B(F)}(y) \simeq \mp \sqrt{\frac{\pi}{2}} y^{3/2} e^{-y} \left(1 +
\frac{15}{8} \frac{1}{y} \right).
\label{jbflowT}
\end{equation}
It is interesting to find where the behavior of $J_{B(F)}$ at low temperature matches
its high-temperature expression. We find that the transition between the low- and
high-temperature approximations occurs at $y \simeq 2.25$ for the
bosonic thermal integral and at $y\simeq 1.85$ for the fermionic
thermal integral. This is sometimes more useful for the numerical analysis
than using the  exact expressions (\ref{jb}) and (\ref{jf}). More
explicitly, we find that a simple interpolation of the two regimes and
a truncation in the high-temperature series in Eqs.~(\ref{jbexp}) and
(\ref{jfexp}), result in the following expressions:

\begin{eqnarray}
J_B(y) & \simeq & \left[-\frac{\pi^4}{45}+
  \frac{\pi^2}{12}y^2-\frac{\pi}{6}
  \left(y^2\right)^{3/2}-\frac{1}{32} y^4\ln\frac{y^2}{a_b} \right.
\nonumber \\
& - & \left.
  \frac{\zeta(3)}{128 \pi^2} y^6\right] \theta(2.25-y)  \nonumber
\\ && - \sqrt{\frac{\pi}{2}} y^{3/2} e^{-y} \left(1 + \frac{15}{8}
\frac{1}{y} \right) \theta(y-2.25), 
\end{eqnarray}
and 

\begin{eqnarray}
J_F(y) & \simeq & \left[
  \frac{7\pi^4}{360}- \frac{\pi^2}{24}y^2-\frac{y^4}{32} \ln
  \frac{y^2}{a_f} \right] \theta(1.85-y) \nonumber \\
&+& \sqrt{\frac{\pi}{2}}
y^{3/2} e^{-y} \left(1 + \frac{15}{8} \frac{1}{y} \right)
\theta(y-1.85),
\end{eqnarray}
which provide an excellent coverage of the exact integrals (\ref{jb}) and
(\ref{jf}), respectively, in the whole region of high and low temperatures. 
It is crucial in the present work to pay particular attention to the specific 
approximation to be used in the
effective potential, due to the large
disparateness of the mass scales that the model has.  At a given temperature,
some particles may acquire a mass that is below the temperature and
others might have a mass above the temperature, so we have different
contributions to the effective potential at different
temperatures. This is particularly important when we investigate the
behavior of the effective potential in between the first transition
and the final  electroweak phase transition in the model.

In the next section we will present the results for the phase
transition pattern in the  3-3-1 model for the different choices of the sets
of parameters explained in Sec.~\ref{scalarspectrum}.


\section{Phase transition pattern in the 3-3-1 model}
\label{PTpattern}

We can clearly identify the two transitions in the model as the
temperature is lowered from values $T \gg v_{\chi_0}$ to values below
the electroweak scale, $T \ll v_W$.  {}First, the higher symmetry
$SU(3)_L \times  U(1)_X$ group is broken down to  the electroweak one,
$SU(2)_L\times U(1)_Y$, at a temperature below the 3-3-1 scale
$v_{\chi_0}$.  Then follows the usual electroweak phase
transition,  $SU(2)_L\times U(1)_Y \to U(1)_{EM}$, at a temperature
slightly below the  Weinberg scale $v_W$.

\begin{center}
\begin{figure}[htb!]
\subfigure[]{\includegraphics[width=7.5cm,height=5.5cm]{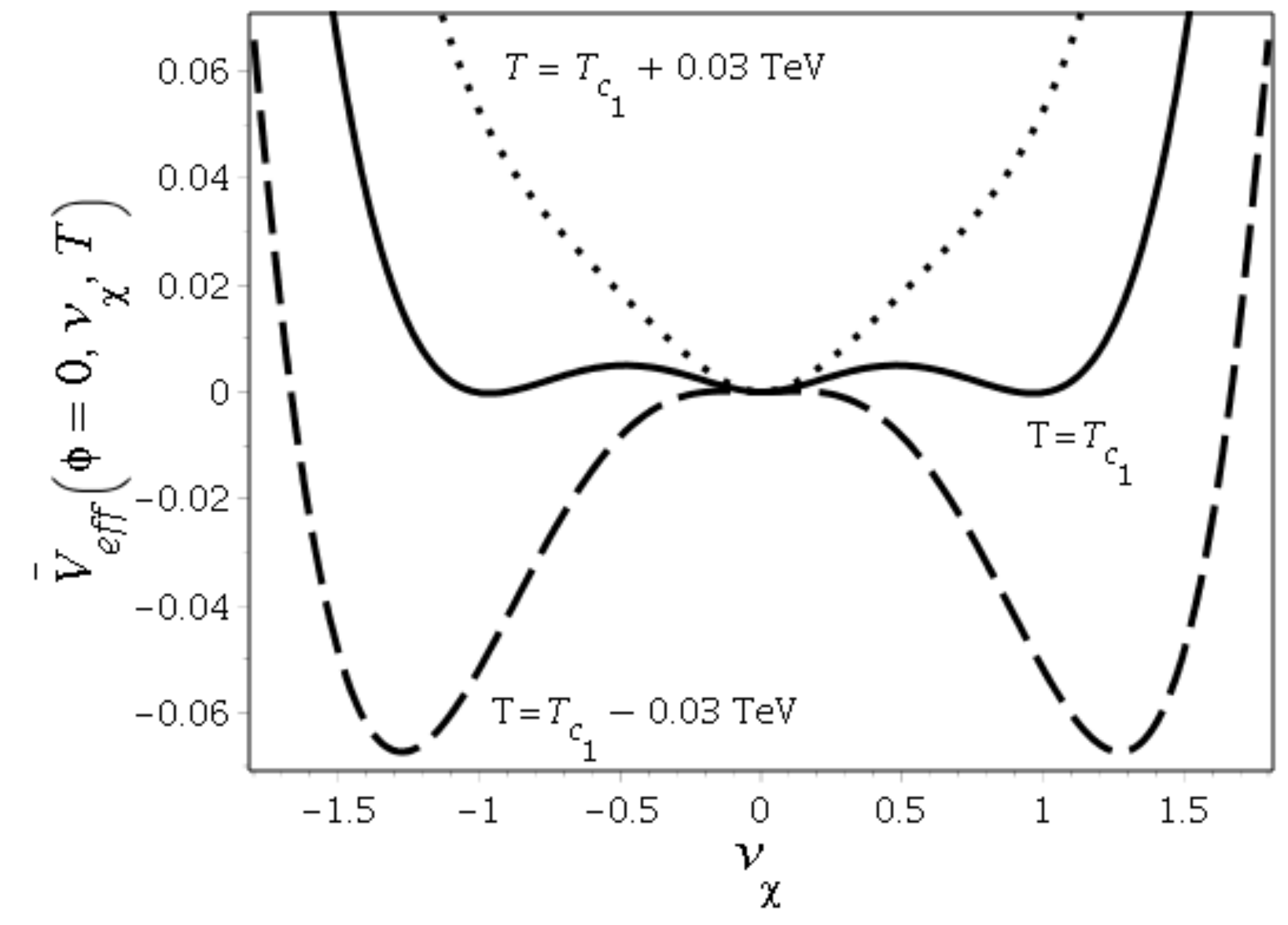}}\hspace{0.5cm}
\subfigure[]{\includegraphics[width=7.5cm,height=5.5cm]{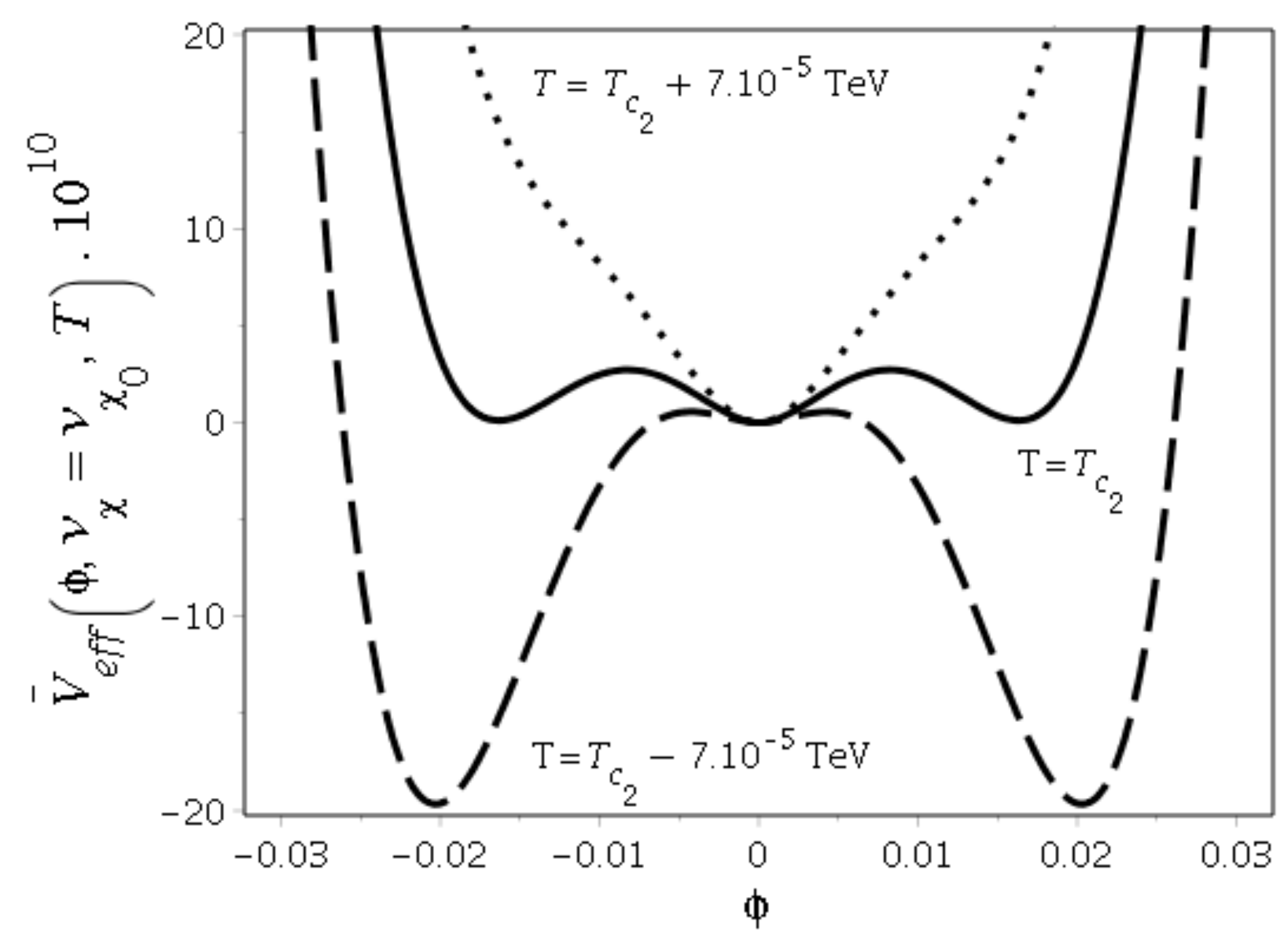}}
\caption{The vacuum subtracted one-loop temperature dependent
  effective potential in the direction of the background scalar field
  $v_\chi$ (at $v_\eta = v_\rho=0$) (panel (a)) and  in the direction
  of the background scalar field $v_\eta = v_\rho$ (panel (b)), expressed in terms
of $\phi = \pm\sqrt{v_\eta^2 + v_\rho^2}$. The temperatures considered are above, at
  and below the critical values. $T_{c_1}$ and $T_{c_2}$
   correspond to the values computed
  at the scale $v_{\chi_0}=3.0$ TeV, for the case of set II of parameters and for
$\tan(\beta)=1$ and whose values are quoted in Table~\ref{tabTc}.  
The potential is in units of
  TeV$^4$ and the background fields are in units of TeV.}
\label{figVeffT}
\end{figure}
\end{center}

As an illustrative example of the phase transition related to the two SSB
in the model, the temperature-dependent potential is shown in {}Fig.~\ref{figVeffT},
considering the parameters of set II, with the couplings at its minimum values,
for $\tan(\beta)=1$ and $v_{\chi_0}=3$ TeV. In {}Fig.~\ref{figVeffT}(a)
we show the effective potential in the direction
of $v_\chi$. It shows the behavior of the effective potential with the
temperature. {}For temperatures above the scale for the electroweak
symmetry breaking, the temperature-dependent values  $v_{\eta}(T)$
and $v_{\rho}(T)$ vanish (or, equivalently, $\phi(T) = \sqrt{v_{\eta}^2(T) + v_{\rho}^2(T)}=0$), 
since the electroweak symmetry is still in the symmetry restored
phase. {}Figure~\ref{figVeffT}(a)  then shows that there is a temperature
$T=T_{c_1}$ for which the potential displays degenerate minima with the one at
the origin. Below this critical temperature the minimum with
non-vanishing background field value becomes the global minimum and
for temperatures slight above  the critical value it is a local
minimum, with the origin being the state of  minimum energy. This
corresponds to a background value for the $\chi$ field, $\langle \chi
\rangle \equiv v_\chi$, that changes discontinuously with the
temperature, jumping from a value $v_\chi =0$ to a nonvanishing value
at the temperature $T=T_{c_1}$.  This is the characteristic of a
first-order phase transition (as opposite to a second-order phase
transition, where the background field changes continuously with the
temperature).
The same behavior as seen in {}Fig.~\ref{figVeffT}(a) is also shown in
{}Fig.~\ref{figVeffT}(b).  In {}Fig.~\ref{figVeffT}(b) we show the effective
potential in the direction of $v_\eta=v_\rho$ (expressed in terms of
$\phi$) for temperatures below the
Weinberg scale $v_W$. {}For these low values of temperature, we have
that  $v_\chi \approx v_{\chi_0}$, i.e., the  thermal expectation
value for the $\chi$ field already approaches its vacuum value
$v_{\chi_0}$. All the heavy particles that make the extra particle
spectrum of the 3-3-1 model acquire masses close to their vacuum
values\footnote{Note that  the heavy particles have a dependence on
  the background fields $v_\eta$ and $v_\rho$, which are, however, zero at and above
  the critical temperature $T_{c_2}$. Thus, all heavy particles, with
  the exception of $Z'$ and the heavy quarks $Q$, which only depend on
  $v_\chi$, will have values close but not exactly at their vacuum
  values.}  and they contribute  little for the effective potential at
this scale. Hence, the particle content dominating the effective
potential at $T \sim T_{c_2} \ll v_{\chi_0}$ is essentially that of
the standard model.  {}Figure~\ref{figVeffT}(b) shows that the final
transition, corresponding to the  standard model one, $SU(2)_L\times
U(1)_Y \to U(1)_{EM}$, happens at a temperature $T=T_{c_2} < v_W$ and
it is of the type of a first-order phase transition.

In Table~\ref{tabTc} we summarize the results for the two phase
transitions in the 3-3-1 model, where we give the value for the critical
temperatures for the two phase transitions in the model. The first one
happens at a temperature $T_{c_1}$ and corresponding to the symmetry
breaking $SU(3)_L \times  U(1)_X \to SU(2)_L\times U(1)_Y$, and the
second transition corresponding to the electroweak symmetry breaking
$SU(2)_L\times U(1)_Y \to U(1)_{EM}$, happens  at the temperature
$T_{c_2}$. Results are shown for the four sets of parameters considered in
this work and for the three different projection angles $\beta$
($v_\eta/v_\rho=\tan (\beta)$).  We also
show the ratio of the background field working as order parameter for
each transition to the temperature at the critical point. As already mentioned,
this is a
useful measure of the "strength" of the phase transition, as usually
considered in the literature~\cite{reviews,Espinosa:1992kf} (for other
alternative forms of characterizing the strength of the transition,
particularly useful for weak first-order phase transitions, see, e.g.,
Refs.~\cite{Gleiser:1992ed,Ramos:1996at}).

\begin{table}[H]
\caption{The critical temperature (in units of TeV) for each of the transitions and the
  ratio of the vacuum expectation value of the relevant field by the
  critical temperature. $T_{c1}$ corresponds to critical temperature
  for the first transition, $SU(3)_L \times  U(1)_X \to SU(2)_L\times
  U(1)_Y$, while $T_{c2}$ corresponds to the one for the second
  transition, $SU(2)_L\times U(1)_Y \to U(1)_{EM}$. The results shown
are for the scale $v_{\chi_0}=3$ TeV.}
\label{tabTc}
\begin{center}
\begin{tabular}{c|c|c|c|c|c}
\hline 
Set  &  $\beta$ &  $T_{c_1}$ & $\langle v_\chi(T_{c_1})\rangle/T_{c_1}$ & $T_{c_2}$  
& $\langle \phi(T_{c_2})\rangle/T_{c_2}$  
\\ \hline   
  &  $30^{\circ}$ & 1.494 & 0.847  & 0.344 & 0.023 
\\  
I  &  $45^{\circ}$ &  1.462 & 1.094  & 0.374 & 0.018  
\\   
  &  $60^{\circ}$ & 1.494 & 0.847 & 0.344  & 0.023  
\\ \hline   
  &  $30^{\circ}$ & 1.701 & 0.800  & 0.205 & 0.080
\\  
 II  &  $45^{\circ}$ &  1.830 & 0.525  & 0.204 & 0.080  
\\   
  &  $60^{\circ}$ & 1.811 & 0.530 & 0.205  & 0.080   
\\ \hline   
  &  $30^{\circ}$ & 1.810  & 0.530  & 0.204 & 0.080  
\\  
 III  &  $45^{\circ}$ & 1.829 & 0.525  & 0.204 & 0.080  
\\   
  &  $60^{\circ}$ & 1.698 & 0.807 & 0.204  & 0.080 
\\ \hline   
  &  $30^{\circ}$ & 1.295 & 1.060  & 0.205 & 0.079  
\\  
 IV  &  $45^{\circ}$ & 0.365 & 8.082  & 0.203 & 0.080   
\\   
  &  $60^{\circ}$ & 1.295 & 1.060 & 0.205  & 0.079    
\\ \hline   
\end{tabular}
\end{center}
\end{table}

In all cases shown in Table~\ref{tabTc} the parameter sets are
taken at their minimal values, shown in {}Fig.~\ref{figsets}.
We have explicitly verified that the values of $T_{c_1}$ specified
correspond to the {\it minimum}  possible critical temperature found
within the range of couplings shown in {}Fig.~\ref{figsets}.
Changing the values of the couplings away from the minimum values
satisfying the Higgs mass constraint always tend to increase the
value of $T_{c_1}$ and, consequently, decrease the ratio $\langle v_\chi(T_{c_1})\rangle/T_{c_1}$. 
The same is true in general for $T_{c_2}$,
except for the case of set I, where we find that decreasing the value of
the coupling $\lambda_4$, $T_{c_2}$ tends to decrease, thus
increasing the ratio $\langle \phi(T_{c_2})\rangle/T_{c_2}$, but the
minimum values of $T_{c_2}$ we have found are still limited by
the minimum values shown in Table~\ref{tabTc}, and 
$\langle \phi(T_{c_2})\rangle/T_{c_2} \lesssim 0.08$.
This then implies that the results for the second transition, corresponding to
 $SU(2)_L\times U(1)_Y \to U(1)_{EM}$, has values for the
ratio $\langle \phi(T_{c_2})\rangle/T_{c_2}$ that are always much smaller than one,
which characterizes a very weak first-order, possibly second-order,
phase transition.  We have explicitly verified that by increasing the scale
$v_{\chi_0}$ it causes very little changes to the second transition.
This is consistent with the fact that the higher is $v_{\chi_0}$, the sooner the
heavier particles decouple from the SM electroweak spectrum. 
We also note that
closer to the transition point it also known that self-energy
corrections to the effective potential can make the transition even
weaker~\cite{Espinosa:1992kf,Carrington:1991hz}.

\begin{center}
\begin{figure}[htb!]
\includegraphics[width=7.5cm,height=5.5cm]{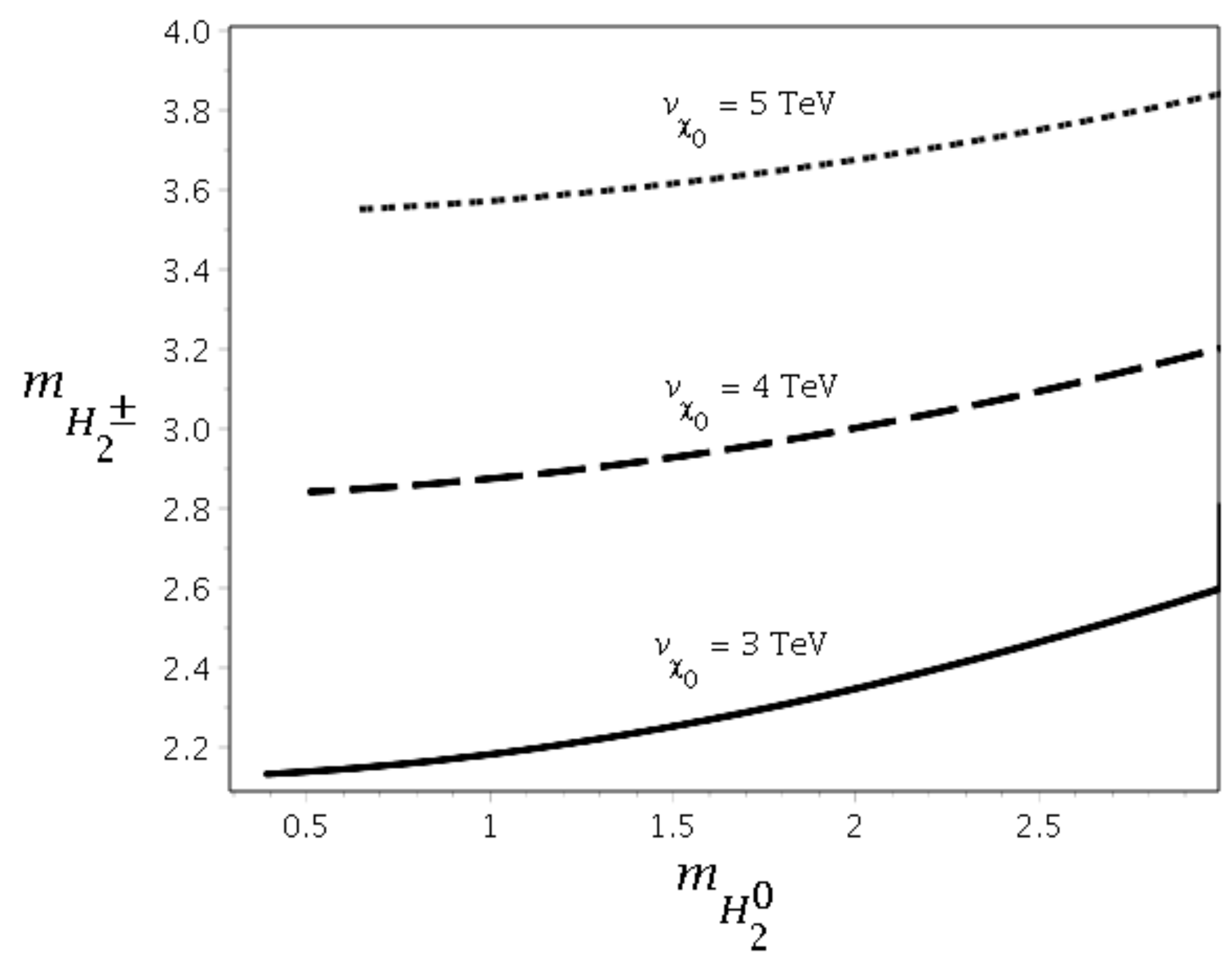}
\caption{The functional dependence for the masses for the scalars
$H_2^0$ and $H_2^\pm$ for the set IV case of parameters and for
$\tan(\beta)=1$. Units are in TeV.}
\label{plotmasses}
\end{figure}
\end{center}

As far as the first transition is concerned,  corresponding to 
$SU(3)_L \times  U(1)_X \to SU(2)_L\times U(1)_Y$, we find that it has a value for
the critical temperature that increases proportionally to the scale $
v_{\chi_0} $, as we would expect on general grounds. The ratio
$\langle v_\chi(T_{c_1})\rangle/T_{c_1} $ tends to be closer to one,
becoming larger when the scale increases. Among the different sets of parameters
we have considered, the most favorable for producing a strong first-order
phase transition,  $\langle v_\chi(T_{c_1})\rangle/T_{c_1}> 1$, is set IV, 
as explicitly noted from the values  shown in Table~\ref{tabTc}.
In particular, the case with $\tan(\beta) =1$, i.e., $v_\eta = v_\rho$,
is the one that is able to produce the strongest transition.
Note, however, looking at the values for the scalar mass spectrum shown in
Table~\ref{spectrum}, this is also the case that leads to the smallest mass for the
Higgs like scalar particle $H_2^0$, with a mass $M_{H_2^0} \simeq 383$ GeV.
Increasing the scale $v_{\chi_0} $ this value of the mass also increases.
{}For example, for $v_{\chi_0}= 5$ TeV, we have $M_{H_2^0} \simeq 616$ GeV
and $T_{c_1} = 605$ GeV, with a ratio $\langle v_\chi(T_{c_1})\rangle/T_{c_1}\simeq 8.12$.
In {}Fig.~\ref{plotmasses} we show the functional dependence for the
masses of the lightest scalars after the standard model Higgs,
i.e., for the scalars $H_2^0$ and $H_2^\pm$ (note that the double
charged scalar $H^{\pm\pm}$ is degenerate in mass with  $H_2^\pm$
when $\tan(\beta)=1$). The value of $T_{c_1}$ tends to increase (and
consequently the ratio  $\langle v_\chi(T_{c_1})\rangle/T_{c_1}$ decreases)
as we move from the smallest values of masses towards the largest values.
{}For example, in the $v_{\chi_0}=3 $ TeV, for $M_{H_2^0} = 1$ TeV, we
find that $T_{c_1}\simeq 1.1$ TeV and  $\langle v_\chi(T_{c_1})\rangle/T_{c_1}
\simeq 1.8$. Typically, we find that for all sets of parameters 
considered, $T_{c_1} \approx M_{H_2^0}$ within around $10\%$. Note that this 
automatically implies a lower bound, $M_{H_2^0} \gtrsim v_W$, since the 
first transition must obviously occur at a temperature above the electroweak one.

\section{Conclusions}
\label{conclusions}

In this work we have studied the symmetry breaking patterns of the
3-3-1 model at finite temperature. Making use of the minimal version
of the model, we have first analyzed its scalar sector, which is
constructed from three scalars in the triplet representation of
$SU(3)_L$ and the most general renormalizable interactions that can be
constructed with these fields. Despite the very large parameter space
of the model, we have made an extensive analysis of the model making
use of four large sets of parameters that give relations between the
scalar couplings. This was done in such a way to maximize the possibility
of finding a strong first-order phase transition on this model, motivated
by its possible role in baryogenesis scenarios in extensions
of the SM.
This allowed us to make a systematic (though far from complete, 
it should be sufficiently representative for our purposes in this work)
investigation the two  symmetry
transitions in the model, the $SU(3)_L \times  U(1)_X \to
SU(2)_L\times U(1)_Y$ and the standard electroweak one,
$SU(2)_L\times U(1)_Y \to U(1)_{EM}$.

On studying the temperature effects on the effective potential at
the one-loop level, and within the approximations used, we have shown
that the model encodes two first-order phase  transitions. The last
one, corresponding to the standard electroweak phase transition,
$SU(2)_L\times U(1)_Y \to U(1)_{EM}$, turns out to be always very weak,
most likely turning into a second-order or a crossover in
practice. {}For the  first transition, corresponding to
$SU(3)_L \times  U(1)_X \to SU(2)_L\times U(1)_Y$,
we find that there are regions of parameters that can favor a strong
first-order phase transition and, in particular, we have found that
the critical temperature in this case is always close to the mass of
the Higgs like scalar $H_2^0$. This indicates that we can use the
estimated value for the mass  $M_{H_2^0}$ as a reasonable estimate
for the temperature of transition $T_{c_1}$.

Our results should be contrasted with some previous analyses
of the phase transition performed in some variants of the 3-3-1 model
done in Refs.~\cite{Phong:2013cfa,Phong:2014ofa}. In the
Ref.~\cite{Phong:2013cfa} the authors have used the so-called reduced
minimal 3-3-1 model, while in Ref.~\cite{Phong:2014ofa} the
economical 3-3-1 model was used. These models differ from the one we have
used in the present work in the sense that they have a reduced number
of couplings in the potential for only two scalar triplets interactions. 
The economical 3-3-1 model has a much richer
leptonic content than the reduced minimal version and both versions 
exclude quarks with exotic electric charges. The
SSB mechanism applied for determining how each ordinary or new gauge boson 
acquires the mass follows the same
road as usual, The Goldstone bosons are identified, but no numerical value 
for the masses of the scalars have been shown.  In
Refs.~\cite{Phong:2013cfa,Phong:2014ofa} the authors find parameter
regimes where  strong first-order phase transitions are found for both
the $SU(3)_L \times  U(1)_X \to  SU(2)_L\times U(1)_Y$ and also for
the  $SU(2)_L\times U(1)_Y \to U(1)_{EM}$ transitions. 
This discrepancy as regards the  strength of the predicted two 
step phase transition in these alternative models with our present results
deserves an interpretation.

We believe that the most important source for the difference between
our results and the previous ones come from the fact that in
Refs.~\cite{Phong:2013cfa,Phong:2014ofa}  it was assumed that only
one field direction would contribute at each transition, e.g., with
$v_\eta =0$ in the first transition along the $v_\chi$ direction and
with $v_\chi =0$ in the second transition along the $v_\eta$
direction. While this is basically true in the first case, where the
temperature is sufficiently high to have $v_{\eta,\rho}=0$, this is not the
case for the second transition. As we have
explained in Sec.~\ref{PTpattern}, in the first transition the electroweak phase would
still be in its symmetry restored phase for temperatures $T \gg
v_W$, thus $v_\eta =v_\rho=0$. However, for the second transition, the temperature is already
low enough, $T < v_W$, so that  $v_\chi \approx v_{\chi_0}$ and all
the heavy particles that make the extra particle spectrum of the 3-3-1
model acquire masses close to their vacuum values decoupling
from the particle spectrum (e.g., their temperature-dependent 
contributions to the effective
potential become all Boltzmann suppressed). The particle content at
these low values of the temperature is then dominated essentially by
that of the standard model. As such, we expect the results not to
differ strongly from the ones known for the phase transition in the
standard model.  That the heavy mass particles of the 3-3-1 model
contributes little at the electroweak  phase transition is confirmed
by the results. We have found in all cases of sets of parameters 
considered here that the
critical temperature $T_{c_2}$, as well the ratio $\langle
v_\eta(T_{c_2})\rangle/T_{c_2} $, are very weakly dependent on the
scale $v_{\chi_0}$, which controls the masses of the heavy particles.

Our results show that using the parameter set IV, in particular for
$\tan(\beta)=1$, can lead to a very low critical temperature for
the first transition.  In particular, 
we have obtained the result that, for all parameters studied, $T_{c_1} \approx M_{H_2^0}$ 
within around $10\%$. This result and the possibility of having 
a critical temperature $T_{c_1}$ not too high above that for the EWPT, $T_{c_2}$,
are deserving of further analysis in the future and may have for this model important 
implications as regards astroparticle physics and cosmology.


\begin{acknowledgements}

R.O.R is partially supported by research grants from Conselho Nacional
de Desenvolvimento Cient\'{\i}fico e Tecnol\'ogico (CNPq)  and
Funda\c{c}\~ao Carlos Chagas Filho de Amparo \`a Pesquisa do Estado do
Rio de Janeiro (FAPERJ).

\end{acknowledgements}


\end{document}